\documentclass[fleqn]{2022AMSE}

\usepackage{graphicx}
\usepackage{amssymb}
\usepackage{amsmath}
\usepackage{booktabs}
\usepackage{adjustbox}

\usepackage{subfig}
\usepackage{caption}
\usepackage{float}      
\usepackage{placeins}   

\usepackage[square,numbers,sort&compress]{natbib}
\makeatletter
\@fleqnfalse
\makeatother

\begin{document}
\setlength{\mathindent}{0cm}

\ensubject{Fluid Dynamics}
\ArticleType{RESEARCH PAPER}
\Year{}
\Vol{}
\DOI{}
\ArtNo{}
\ReceiveDate{???}
\AcceptDate{???}
\OnlineDate{???}

\title{RETO: A Rotary-Enhanced Transformer Operator for High-Fidelity  Prediction of Automotive Aerodynamics}

\author[1,2]{Bojun Zhang}{}
\author[1,2]{Huiyu Yang}{}
\author[1,2]{Yunpeng Wang}{}
\author[3]{\\Yuntian Chen}{}
\author[3,4]{Yuanwei Bin}{}
\author[4]{Rikui Zhang}{}
\author[1,2]{Jianchun Wang}{}\thanks{Corresponding author. E-mail:
\href{mailto:wangjc@sustech.edu.cn}{wangjc@sustech.edu.cn}}

\AuthorMark{Bojun Zhang}

\AuthorCitation{Bojun Zhang, Huiyu Yang, Yunpeng Wang, Yuntian Chen, Yuanwei Bin, Rikui Zhang, Jianchan Wang}

\address[1]{Department of Mechanics and Aerospace Engineering, Southern University of Science and Technology, Shenzhen, 518055, China}
\address[2]{Shenzhen Key Laboratory of Complex Aerospace Flows,  Southern University of Science and Technology,Shenzhen 518055, China}
\address[3]{Ningbo Key Laboratory of Advanced Manufacturing Simulation, Eastern Institute of Technology, Ningbo, 315200, China}
\address[4]{Shenzhen Tenfong Science and Technology Co., Ltd., Shenzhen, 518000, China}

\contributions{Executive Editor: ???}

\abstract
{

{Rapid aerodynamic evaluation is critical for vehicle design, yet neural operators struggle with complex spatial correlations. We propose a rotary-enhanced transformer operator (RETO), with a dual-stage spatial encoding: sinusoidal-cosines for global referencing and rotary positional encodings (RoPE) for relative displacements. RoPE enforces translation invariance and enhances local gradient resolution via unitary rotations. Validated on the ShapeNet and DrivAerML datasets, RETO achieves a relative $L_2$ error of 0.063 on the ShapeNet surface pressure field, outperforming RegDGCNN (0.125) and Transolver (0.075). On the DrivAerML, RETO attains relative $L_2$ errors of 0.089 for surface pressure and 0.097 for velocity, outperforming Transolver (0.116/0.121), Transolver++ (0.127/0.133), and Transolver-3 (0.121/0.136) with relative improvements of 19\%--30\%. Information-theoretic analysis reveals a lower entropy peak (0.35 vs. 0.75 for Transolver at $10^4$ resolution), confirming focused attention that preserves local gradients against global diffusion. The code is available at: https://github.com/bojunZhang-heng/RETO}

}


\setlength{\textheight}{23.6cm}
\thispagestyle{empty}

\maketitle
\setlength{\parindent}{1em}

\begin{multicols}{2}
\section{Introduction}
In modern automotive engineering, aerodynamic optimization is a multi-objective task essential for enhancing fuel efficiency \cite{abinesh2014cfd}, extending the range of electric vehicles \cite{bibra2022global}, and ensuring high-speed handling stability \cite{ang2023aerodynamic}. High-fidelity aerodynamic analysis necessitates the simultaneous evaluation of two pivotal physical fields: the volumetric velocity field, which reveals wake dynamics \cite{islam2017detailed,zhang2020experimental} and cooling flow patterns \cite{2013Interference}; the surface pressure field, which dictates the distribution of lift and drag forces \cite{2023Automotive,huang2024neural}. The synergy of these coupled physical fields constitutes a holistic representation of the vehicle aerodynamic signature, providing the essential mechanistic insights required to navigate the complex trade-offs between competing design objectives.

For decades, computational fluid dynamics (CFD) methods \cite{anderson1995computational,ferziger2019computational} are employed to resolve complex flow fields, fundamentally relying on the numerical solution of the Navier-Stokes  equations \cite{pope2001turbulent}. Various classical numerical frameworks have been established, including the finite difference method \cite{smith1985numerical}, the finite element method \cite{hughes2003finite}, the finite volume method \cite{eymard2000finite}, and the lattice Boltzmann method \cite{chen1998lattice}. Within these discretization frameworks, the selection of an appropriate turbulence modeling paradigm is typically dictated by the specific engineering objective fidelity. For instance, Reynolds-averaged Navier-Stokes equations (RANS) \cite{aultman2022evaluation,wilcox2008formulation,menter1992improved} approaches remain the industrial workhorse for evaluating steady-state aerodynamic loads, including lift and drag, due to their relatively high computational efficiency. In contrast, resolving the transient, high-frequency vortex shedding and pressure fluctuations essential for aeroacoustic assessments necessitates more computationally intensive large eddy simulation (LES) \cite{lee2022development,piomelli1996large}. However, regardless of the numerical scheme, resolving complex automotive flow fields remains computationally expensive, often requiring hours or even days for a single design evaluation. This significant cost impedes rapid design iteration and optimization. 

Recently, data-driven deep learning methods\cite{shan2024modeling,bai2025towards}, have introduced new paradigms for the rapid solution of partial differential equations (PDEs) \cite{lu2021learning,li2020fourier,hao2025large}. In data-driven aerodynamic assessment, convolutional neural networks (CNNs) \cite{garcia2023cnn,liu2026reconstructing} have been widely adopted for vehicle surface field prediction due to their high efficiency when geometric information is represented in structured forms, as exemplified by benchmarks such as DrivAerNet++ \cite{elrefaie2024drivaernet++}.
However, the reliance on local convolutional operators acting on such structured representations limits their effectiveness for volumetric velocity field prediction, where capturing large-scale wake interactions and handling irregular spatial distributions are essential.
In contrast, Transformer architectures \cite{vaswani2017attention}, built upon self-attention mechanisms, provide a more flexible framework for modeling long-range dependencies and unstructured data \cite{yang2025spatially}. By treating discrete flow samples as independent tokens \cite{li2022transformer}, Transformers can operate directly on point clouds \cite{qi2017pointnet++}, enabling global information exchange across the entire spatial domain. This capability endows Transformer-based models with enhanced robustness and scalability in the prediction of three-dimensional physical fields \cite{alkin2025ab,Liu2026EMOS,Liu2026AeroAgent}, including velocity and pressure.

Here, we briefly summarize some related works on surrogate models for rapid aerodynamic assessments.

\textbf{Neural operators}. Neural operators construct mesh-invariant mappings from infinite-dimensional parameter spaces to solution spaces \cite{lu2021learning, kovachki2023neural}, enabling efficient approximation of solution operators for partial differential equations,such as the Navier-Stokes equations. Lu et al. proposed the  deep operator network (DeepONet) \cite{lu2019deeponet}, a dual-network architecture where branch and trunk networks separately encode input functions and output locations. Li et al. introduced the Fourier neural operator (FNO) \cite{li2020fourier} , which leverages fast Fourier transforms for efficient spectral integration, achieving fast operator learning while being constrained to regular grids. Building on this work, Tran et al.proposed factorized Fourier neural operator (F-FNO) \cite{tran2021factorized} which consists of separable fourier representation and improved residual connections, reducing the model complexity and allowing it to scale to deeper network. In summary, the neural operators provide a computationally efficient framework for approximating solution of PDE and thereby enable rapid inference of flow fields; however, their predictive accuracy and stability deteriorate when exposed to highly irregular or non-smooth inputs, which limits their robustness in complex industrial aerodynamic problems.

\textbf{Convolutional neural networks-based method}. Data-driven models based on large amounts of data with direct output of flow field characteristics have been well studied, and CNN is one of the popular approaches \cite{krizhevsky2012imagenet,lecun2002gradient}. Bhatnagar et al. proposed a CNN-based surrogate model \cite{bhatnagar2019prediction} which predicted velocity and pressure fields under unseen flow conditions. Wang et al. proposed edge convolution (EdgeConv) \cite{wang2019dynamic}, which captures semantic correlations by dynamically reconstructing graph topologies in high-dimensional feature spaces. To address the fidelity degradation of CNNs in 3D wake and near-wall regions, Chen et al. proposed an enhanced estimation model incorporating a slice-weighted loss function and fluid continuity constraints to refine volumetric flow predictions \cite{chen20213d}. Garcia et al. employed signed distance functions and a non-uniform CNN mesh to predict velocity and pressure fields around heavy vehicles \cite{garcia2023cnn}. Elrefaie et al. proposed dynamic graph convolutional neural network for regression (RegDGCNN) \cite{elrefaie2025drivaernet} which presents a dynamic graph convolutional model that operates directly on high-resolution 3D meshes to enable efficient and accurate aerodynamic drag prediction across diverse geometries. Although convolutional operators are effective at extracting the physical information of surface fields, they exhibit significant limitations when it comes to predicting fully three-dimensional spatial flow physics.

\textbf{Transformer-based methods}. Originally developed for natural language processing, the Transformer architecture \cite{vaswani2017attention} has established itself as a fundamental component of large language models \cite{touvron2023llama}. Its attention-based mechanism for modeling long-range dependencies has recently been adapted to the solution of PDE \cite{wu2022flowformer, jiang2023transcfd}, enabling improved representation of global physical interactions compared with local convolution-based approaches. Transolver \cite{wu2024transolver} leverages physics-attention to learn intrinsic physical states, demonstrating competitive performance with strong scalability and generalization in large-scale industrial CFD simulations. Building upon Transolver, Transolver++ \cite{luo2025transolver++} increases the single-GPU input capacity to million-scale points for the first time and is capable of continuously scaling input size in linear complexity by increasing GPUs. Transolver-3 \cite{zhou2026transolver} enables high-fidelity PDE solving on industrial-scale meshes by introducing optimized Physics-Attention (via operation reordering and tiling), along with geometry amortized training and decoupled inference with state caching. GeoTransolver \cite{adams2025geotransolver} extends Transolver by integrating geometry-aware latent encodings to couple physics-aware self-attention on learned slices with cross-attention to persistent multi-scale geometry and global context from ball queries. GeoFormer \cite{gu5601950geoformer} introduces a mesh-free geometry-to-flow alignment framework that utilizes a spatial cross-attention mechanism to directly map non-watertight CAD geometries to surface and volumetric flow fields while maintaining $O(N)$ linear complexity. HMT-PF \cite{du2025spatiotemporal} integrates a hybrid Mamba-Transformer backbone with a physics-informed fine-tuning mechanism that leverages self-supervised learning to effectively reduce physical equation discrepancies while preserving essential spatiotemporal field characteristics. Multi-scale patch transformer (MSPT) \cite{curvo2025mspt} introduces a parallelized multi-scale attention mechanism that, combined with ball tree spatial partitioning and hierarchical pooling, simultaneously captures fine-grained local interactions and long-range global dependencies within a unified framework, enabling scalable and accurate physical modeling on million-scale point clouds with a single GPU. Spatially-aware transformer operator (SATO) \cite{yang2025spatially} integrates global and local spatial correlation modeling through complementary attention mechanisms. DragSolver \cite{liu2025dragsolver} utilizes a multi-scale Transformer with surface-guided gating and Monte Carlo dropout to achieve robust, real-time $C_d$ estimation on complex vehicle geometries. Overall, transformer-based operator learners are effective for irregular, high-dimensional data, but may still struggle to capture fine-scale local features due to their global attention mechanisms.

\textbf{Contributions of This Work}. This paper proposes a novel rotary-enhanced transformer operator (RETO). The framework constructs a dual-stage spatial awareness mechanism, utilizing sinusoidal-cosine encodings to establish a global geometric reference, while integrating rotary positional encodings \cite{su2024roformer} to modulate the query and key interactions within the attention mechanism. Although originally developed for Large Language Models, the effectiveness of such encoding has not yet been established for fluid dynamics, particularly in the context of complex vehicle geometries. Evaluations on the industrial-grade DrivAerML dataset \cite{ashton2024drivaerml} demonstrate that RETO achieves significant breakthroughs in predicting volumetric velocity and surface pressure fileds. Compared to the popular Transolver \cite{wu2024transolver}, RETO reduces the relative $L_2$ error by 19\% for the velocity field and 29\% for the pressure field. This study establishes a robust and high-precision mathematical tool for real-time automotive aerodynamic analysis.

\section{Methodology}
In this section, we introduce RETO, a framework designed to resolve high-fidelity aerodynamic fields over complex 3D vehicle geometries via a dual-stage spatial awareness mechanism. RETO encodes relative spatial displacements as rotation matrices within the complex domain, enabling
multi-scale feature extraction that enhances both local and global feature representations.

\subsection{Problem definition}
In automotive aerodynamic simulations \cite{ahmed1984some}, the flow field is typically characterized as unsteady, incompressible flow of a Newtonian fluid, wherein the fluid density is treated as a constant. The physical process is approximated by the RANS equations \cite{wilcox1998turbulence, pope2001turbulent,menter2003ten}, which can be written as:

\begin{equation}
\frac{\partial U_i}{\partial x_i} = 0 ,
\label{eq:continuity}
\end{equation}

\begin{equation}
\frac{\partial U_i}{\partial t} + \frac{\partial (U_j U_i)}{\partial x_j}
= - \frac{\partial p^{*}}{\partial x_i}
+ \frac{\partial}{\partial x_j}
\left[
\left( \nu + \nu_t \right)
\left(
\frac{\partial U_i}{\partial x_j}
+ \frac{\partial U_j}{\partial x_i}
\right)
\right] ,
\label{eq:momentum}
\end{equation}
where Eq.~\eqref{eq:continuity} describes the conservation of mass and Eq.~\eqref{eq:momentum} the conservation of momentum.The flow state is characterized by the velocity field $U_i$ and the density-normalized pressure $p^*$. While $\nu$ accounts for the molecular kinematic viscosity of fluid, the effective viscosity of the system is augmented by the turbulent eddy viscosity $\nu_t$. This quantity, $\nu_t$, is a modeled parameter derived from the Boussinesq linear eddy viscosity assumption \cite{schmitt2007boussinesq}. By incorporating $\nu_t$, the RANS-based formulation effectively captures the macro-scale dissipative effects of turbulence within the automotive wake without explicitly resolving the small scales \cite{swanson2021solving}.

The flow turbulence characteristics are defined by the Reynolds number, which represents the ratio of inertial forces to viscous forces:

\begin{equation}
Re = \frac{U_{\infty} L_c}{\nu} \,,
\end{equation}
where $U_{\infty}$ denotes the free-stream velocity, $L_c$ is the characteristic length of the vehicle (typically its total length), and $\nu$ is the molecular kinematic viscosity.

In this study, we focus on high-fidelity aerodynamic analysis derived from transient simulations. Our primary objective is to develop a data-driven surrogate model that captures the mean flow characteristics by learning the mapping from a vehicle surface geometry to its corresponding time-averaged physical fields. Specifically, the DrivAerML data is generated using delayed detached eddy simulation (DDES) \cite{spalart2009detached}, with the target outputs obtained through time-averaging the unsteady solutions to represent the stabilized mean state. The proposed network directly maps the vehicle geometry, represented as a point cloud, to both the surface pressure and the volumetric velocity fields. While surface pressure is fundamental for calculating aerodynamic forces idrag and lift, the spatial velocity field provides critical insights into complex wake structures and flow separation. Compared to conventional DDES, which requires massive computational resources to resolve transient eddies over extended physical time scales, the data-driven approach bypasses the need for iterative solvers and dense volumetric meshing during inference, enabling near-instantaneous prediction of high-fidelity mean flow distributions directly from the input geometry.

\subsection{Standard Transformer Architecture}
The core of the Transformer architecture is the self-attention mechanism \cite{2014Neural}, which in the context of neural operators, is reinterpreted as a method to efficiently capture the complex, non-local dependencies inherent in fluid dynamics. As illustrated in Fig.\ref{fig:Transformer_Architecture}, the attention module takes query ($Q$), key ($K$), and value ($V$) representations as input and computes attention weights through a scaled dot-product operation. These weights are then used for weighted aggregation of the value features, followed by an output projection. This mechanism allows the attention weights to scale the information within the value representations, enabling each spatial point to interact with every other point in the computational domain. Unlike local convolutional operators, this mechanism effectively constructs a global receptive field essential for resolving long-range fluid interactions.

\begin{center}
    \begin{minipage}{0.15\textwidth}
        \centering
        \includegraphics[width=\linewidth]{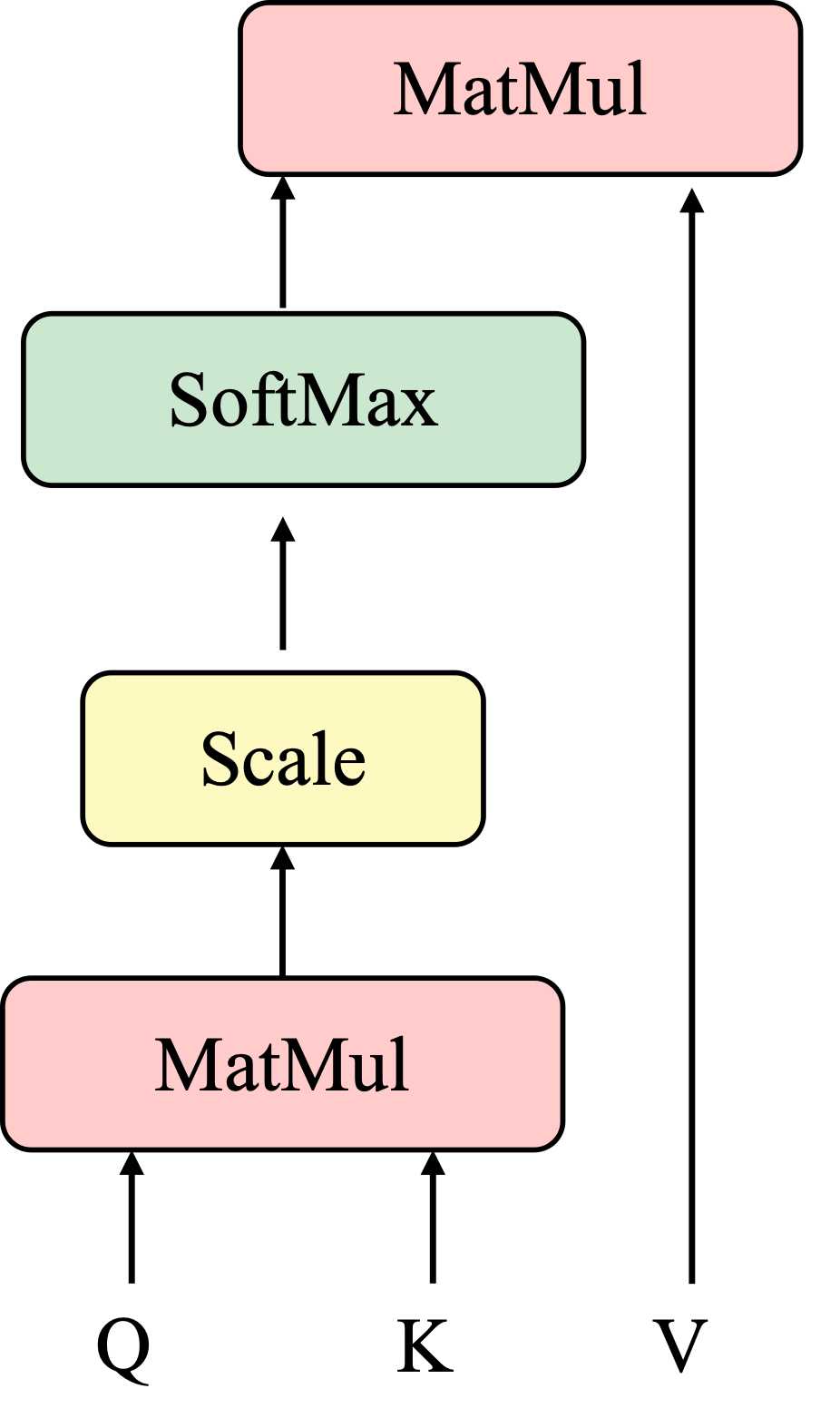}
        \label{fig:multihead_attention}
        (a) Self attention
    \end{minipage}
    \hspace{0.05\textwidth}
    \begin{minipage}{0.2\textwidth}
        \centering
        \includegraphics[width=\linewidth]{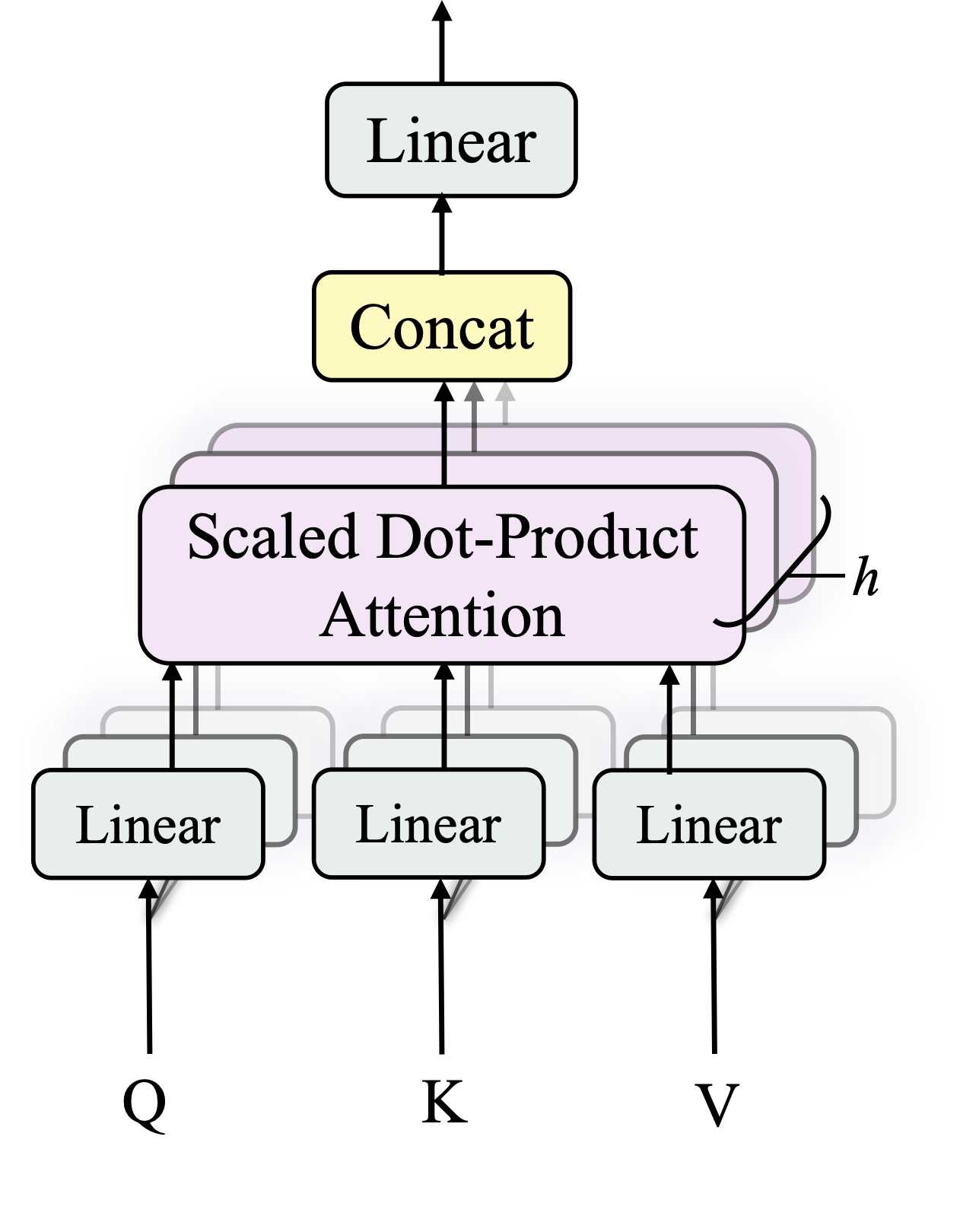}
        \label{fig:dot_plot}
        (b) Multi-Head Attention
    \end{minipage}

    \captionof{figure}{Transformer Architecture.}
    \label{fig:Transformer_Architecture}
\end{center}

Let $X \in \mathbb{R}^{N \times D}$ represent a sequence of $N$ spatial points encoded in a $D$-dimensional feature space. The input is first projected into three distinct latent spaces—$Q$, $K$, and $V$—via learnable weight matrices $W^Q, W^K, W^V \in R^{D \times D_k}$ \cite{vaswani2017attention}:

\begin{equation}
Q = XW^Q,\qquad 
K = XW^K, \qquad
V = XW^V,
\end{equation}
where $D_{k}$ is the dimension of key. The scaled dot-product attention, as illustrated in Fig. \ref{fig:Transformer_Architecture} (a), is then formulated as \cite{2015Effective}:

\begin{equation}
\text{Attention}(Q, K, V) = \text{softmax}\left(\frac{QK^T}{\sqrt{D_k}}\right) V\,.
\end{equation}

To capture diverse physical features across multiple scales, multi-head attention (MHA) is employed, as illustrated in Fig. \ref{fig:Transformer_Architecture} (b). A total of $h = 8$ parallel attention heads are utilized, each focusing on a distinct subspace:

\begin{equation}
\text{MHA}(Q, K, V) = \text{Concat}(\text{head}_1, \dots, \text{head}_h)W^O,
\end{equation}
where $\text{head}_i = \text{Attention}(QW_i^Q, KW_i^K,VW_i^V)$, $W^O \in \mathbb{R}^{hD_v \times D_{\text{model}}}$. While this mechanism provides a powerful framework for modeling unstructured flow data, the standard self-attention is inherently permutation-invariant. In the absence of an rotary positional encoding \cite{su2024roformer}, the model treats the spatial point cloud as an unordered set, neglecting the critical relative distances influence. This limitation motivates the integration of specialized spatial encodings, as discussed in the following section.

\subsection{RETO Architecture}

\begin{center}
  \includegraphics[width=0.5\textwidth]{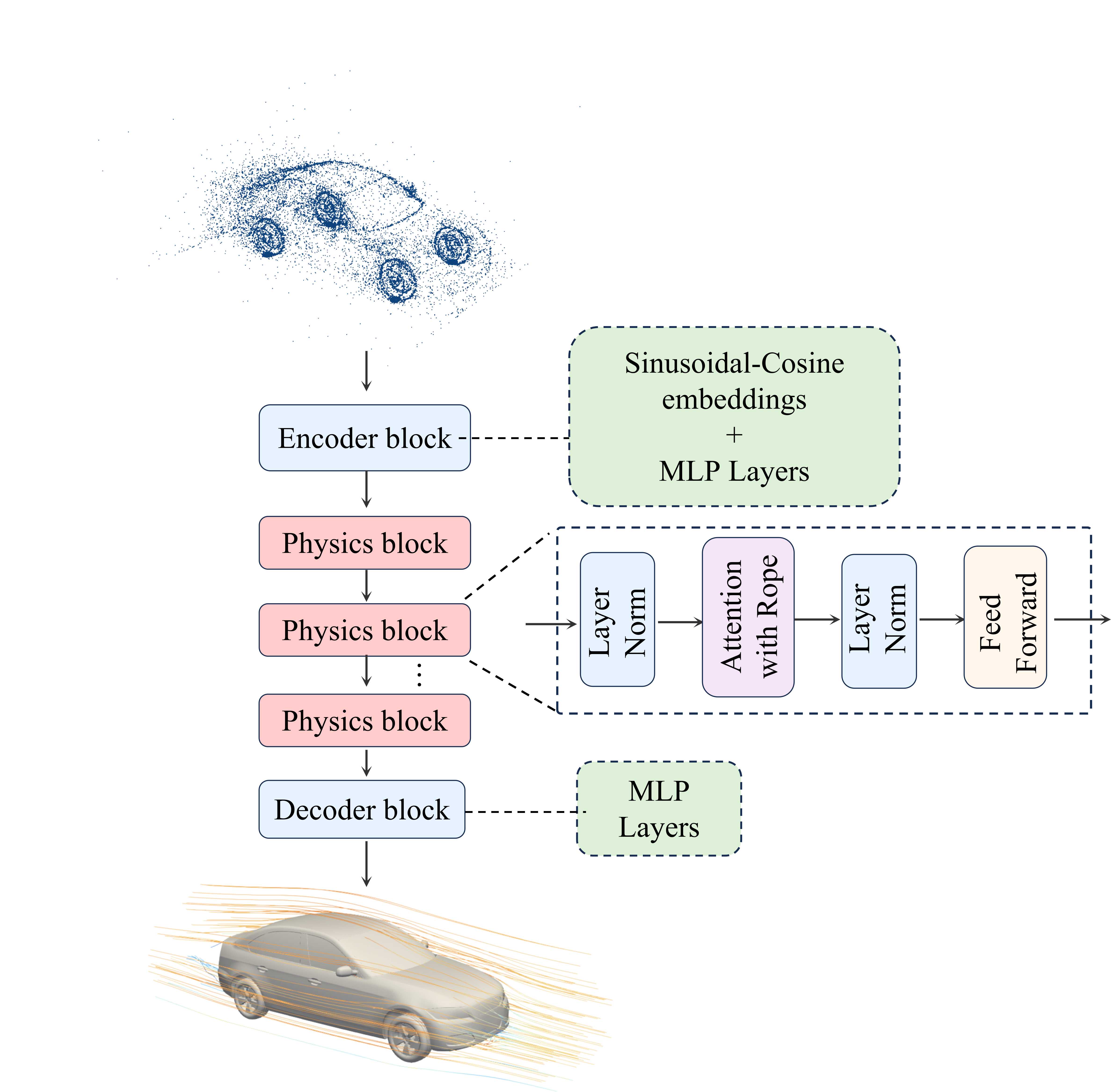}
  \captionof{figure}{RETO Architecture}
  \label{fig:RETO-Architecture}
\end{center}


The proposed RETO architecture is organized into three main components, namely an encoder, a sequence of $T$ stacked physics blocks (where $T=5$ in this study), and a decoder, as illustrated in Fig.~\ref{fig:RETO-Architecture}. The encoder and decoder form the input–output interfaces of the RETO architecture. Specifically, the encoder applies a continuous sine–cosine positional encoding \cite{vaswani2017attention} to expand each input spatial coordinate into a multi-scale spectral representation. The input to the encoder is a tensor of shape $(B, N, C)$, where $B$ is the batch size, $N$ the number of downsampled points, and $C$ the spatial dimension (typically $C=3$). For each point $\boldsymbol{x} \in \mathbb{R}^C$, its coordinate components $x_i$ are mapped to a higher-dimensional latent space via a frequency mapping operator $\boldsymbol{\omega}$. The components of this operator are constructed as a geometric progression based on a maximum wavelength $\lambda = 10,000$ and the target dimensionality per axis $m$:
\begin{equation}
\omega_n = \frac{1}{\lambda^{\frac{2n}{m}}}, \quad n \in {0, 1, \dots, \frac{m}{2} - 1} \,,
\end{equation}
where the index $2n$ represents an arithmetic progression with a constant interval of 2. Let $\boldsymbol{\omega}$ be the column vector of frequencies. The spectral encoding function $\gamma(x_i)$ is then computed via the matrix multiplication of the spatial coordinate and the frequency operator:
\begin{equation}
\gamma(x_i) = \left[ \sin(x_i \boldsymbol{\omega}^T), \cos(x_i \boldsymbol{\omega}^T) \right] \in \mathbb{R}^{m} \,.
\end{equation}

The final latent representation produced by the encoder, $X_{enc}$, is obtained by concatenating the encoded features of all three spatial dimensions:
\begin{equation}
X_{enc} = \text{Concat}(\gamma(x), \gamma(y), \gamma(z)) \in \mathbb{R}^{3m} \,,
\end{equation}
where $3m=D$. For a batch of $N$ downsampled points, the encoded tensor is of shape $(B, N, D)$. This transformation effectively maps the low-dimensional spatial inputs into a high-dimensional feature space, enabling the subsequent physics blocks to resolve fine-grained aerodynamic structures across multiple scales.

To facilitate the integration of these high-dimensional features into the downstream physics blocks, a single-layer perceptron multilayer perceptron (MLP) is used to project the frequency vectors into a compact latent manifold. The MLP has input, hidden, and output dimensions of 256, 512, and 256, respectively. Conversely, a symmetric decoder MLP maps the 256-dimensional latent to an output of shape $(B, N, C_{\text{out}})$, where $C_{\text{out}}=1$ for pressure and $C_{\text{out}}=3$ for velocity.

In the physics block, to enhance the spatial feature extraction capability of the operator learning process, we incorporate rotary positional embedding \cite{su2024roformer}. The core principle involves mapping the feature vectors into a complex domain and applying rotations to capture relative spatial relationships.
Given a $D_h$-dimensional query vector $\boldsymbol{q} \in \mathbb{R}^{D_h}$ (where $D_h$ is the head dimension) at a spatial position $\boldsymbol{x} \in \mathbb{R}^3$, we first interpret $\boldsymbol{q}$ as a set of $D_h/2$ complex numbers. The $n$-th complex pair $(n \in \{1, \dots, D_h/2\})$ is defined as:
\begin{equation}
q^{(n)} = q_{2(n-1)} + i q_{2n}\,,
\end{equation}
where $i$ is the imaginary unit. To inject the spatial information of coordinate $\boldsymbol{x}$, the rotary transformation $\mathcal{R}$ is applied by multiplying the complex vector by a phase shift \cite{su2024roformer}:
\begin{equation}
\tilde{q}^{(n)} = q^{(n)} e^{i \theta_n(\boldsymbol{x})}, \quad \text{with} \quad \theta_n(\boldsymbol{x}) = \sum_{p=1}^3 x_p \cdot \omega_{n,p} \,,
\end{equation}
where $\boldsymbol{x} = (x, y, z)$ represents the Cartesian coordinates of the mesh node, and ${\omega}_n$ is the frequency vector associated with the $n$-th dimension.This transformation can be efficiently implemented as a block-diagonal rotation matrix $\boldsymbol{R}_{\Theta, \boldsymbol{x}}$:

\begin{equation}
\boldsymbol{R}_{\Theta,\boldsymbol{x}} =
\begin{pmatrix}
\cos\theta_1 & -\sin\theta_1 & 0 & \cdots & 0 & 0\\
\sin\theta_1 &  \cos\theta_1 & 0 & \cdots & 0 & 0\\
0 & 0 & \ddots & \ddots & \vdots & \vdots\\
\vdots & \vdots & \ddots & \ddots & 0 & 0\\
0 & 0 & \cdots & 0 & \cos\theta_{d/2} & -\sin\theta_{d/2}\\
0 & 0 & \cdots & 0 & \sin\theta_{d/2} &  \cos\theta_{d/2} 
\end{pmatrix} \,,
\end{equation}
the modified query and key vectors are then $\boldsymbol{\tilde{q}}_i = \boldsymbol{R}_{\Theta, \boldsymbol{x}_i} \boldsymbol{q}_i$ and $\boldsymbol{\tilde{k}}_j = \boldsymbol{R}_{\Theta, \boldsymbol{x}_j} \boldsymbol{k}_j$.

The fundamental advantage of this rotary mechanism lies in the preservation of relative spatial relationships during the self-attention operation. The score function, defined by the inner product, satisfies the following property:

\begin{equation}
\phi_{ij}^{(n)} = \theta_n(\boldsymbol{x}_i) - \theta_n(\boldsymbol{x}_j)\,,
\end{equation}

\begin{equation}
\begin{split}
\left\langle \boldsymbol{R}_{\Theta, \boldsymbol{x}_i} \boldsymbol{q}_i,\ 
\boldsymbol{R}_{\Theta, \boldsymbol{x}_j} \boldsymbol{k}_j \right\rangle
&= \mathrm{Re} \Bigg[
\sum_{n=1}^{D_h/2}
q_i^{(n)} \left(k_j^{(n)}\right)^*
e^{i \phi_{ij}^{(n)}}
\Bigg]\,,
\end{split}
\end{equation}
where $(k_j^{(n)})^*$ denotes the complex conjugate. This identity explicitly demonstrates that the attention scores are functions of the relative displacement $\Delta \boldsymbol{x} = \boldsymbol{x}_i - \boldsymbol{x}_j$:

\begin{equation}
\theta_n(\boldsymbol{x}_i) - \theta_n(\boldsymbol{x}_j) = (\boldsymbol{x}_i - \boldsymbol{ x}_j) \cdot \boldsymbol{\omega}_n \,.
\end{equation}

Physically, this implies that point interactions depend on relative distance: larger spatial separation increases phase differences, attenuating long-range correlations via oscillatory decay. This mirrors the aerodynamic reality where local disturbances (e.g., side mirrors) diminish with distance, balancing local focus and global awareness.

\section{Numerical experiments}

To comprehensively evaluate the robustness and generalization capability of the proposed RETO, we conduct numerical experiments on two datasets \textbf{ShapeNet} \cite{chang2015shapenet} and \textbf{DrivAerML} \cite{ashton2024drivaerml}, benchmarking it against several state of art methods.

\subsection{Implementation details}

The proposed RETO is implemented using the PyTorch deep learning framework. All numerical experiments, including model training and inference, are conducted on a high-performance computing workstation equipped with an NVIDIA Tesla V100-SXM2-32GB.

\textbf{Data pre-processing.} To ensure numerical stability and accelerate the convergence of the training process, the target physical quantities including velocity components and pressure are standardized using the Z-score normalization method \cite{wasserman2004all}. This involves scaling the data based on the mean and standard deviation calculated from the training set:
\begin{equation}
\hat{u} = \frac{u - \mu_u}{\sigma_u} \,,
\end{equation}
where $u$ represents the raw physical value, and $\mu_u$ and $\sigma_u$ are the corresponding mean and standard deviation, respectively.

\textbf{Model Configuration and Hyperparameters.} During the training phase, the model parameters are optimized using the Adam optimizer \cite{2014Adam}. To facilitate stable convergence, the learning rate is regulated by a StepLR scheduler with an initial value of $1 \times 10^{-3}$. The detail is summarized in Table~\ref{tab:training_config}. Epochs is the number of training iterations over the full dataset. Batch Size denotes the number of samples per update. Initial Lr is the starting learning rate. Optimizer specifies the algorithm used for parameter updates, e.g., Adam. Lr scheduler defines how the learning rate changes during training, e.g., StepLR reduces it by a factor of 0.5 every 50 epochs. The architectural hyperparameters of the proposed model are detailed in Table~\ref{tab:hyperparameters}, where $T, H, D$, and $M$ denote the number of physics blocks, attention heads, latent feature dimensionality, and the Transolver-inherited slice number, respectively.

\begin{table*}
\centering
\caption{The training configurations of RETO in two datasets.}
\begin{tabular*}{\textwidth}{@{\extracolsep{\fill}} l c c c c c}
\hline
Datasets  & Epochs & Batch Size & Initial Lr & Optimizer & Lr scheduler \\
\hline
ShapeNet  & 150    & 1    & 0.001    & Adam & StepLR(factor=0.5, step\_size=50) \\
DrivAerML & 250    & 1    & 0.001    & Adam & StepLR(factor=0.5, step\_size=50) \\
\hline
\end{tabular*}
\label{tab:training_config}
\end{table*}

\begin{table*}
\centering
\caption{The model hyperparameters of RETO for two models.}
\begin{tabular*}{\textwidth}{@{\extracolsep{\fill}} l c c c c c}
\hline
Model       & k    & Layers $T$ & Heads $H$ & Dim $D$ & $M$ \\
\hline
RegDGCNN    & 40   & /    & /    & 1024    & /
\\
AB-UBT      & /    & 5    & 8    & 256     & /
\\
Transolver  & /    & 5    & 8    & 256     & 64  \\
RETO        & /    & 5    & 8    & 256     & /   \\
\hline
\end{tabular*}
\label{tab:hyperparameters}
\end{table*}

\textbf{Evaluation metrics.} To quantitatively assess the predictive accuracy and reconstruction fidelity of the proposed model, the relative $L_2$ error is employed as the primary evaluation metric. This metric provides a normalized measure of the discrepancy between the predicted aerodynamic fields and the CFD ground truth, ensuring that the error assessment is independent of the physical quantities.
The relative $L_2$ error is defined as follows \cite{2006Numerical}:
\begin{equation}
{L_2} = \frac{| \boldsymbol{\hat{q}} - \boldsymbol{q} |_2}{| \boldsymbol{q}|_2} = \sqrt{\frac{\sum_{i=1}^N (\hat{q}_i - q_i)^2}{\sum_{i=1}^N q_i^2}} \,,
\end{equation}
where $\boldsymbol{\hat{q}}$ and $\boldsymbol{q}$ represent the predicted physical field and the corresponding ground truth obtained from high-fidelity CFD simulations, respectively; $q_i$ and $\hat{q}_i$ denote the values at the $i$-th spatial point; $N$ is the total number of points within the computational domain or on the vehicle surface.

\subsection{ShapeNet}
The ShapeNet Car dataset \cite{chang2015shapenet} serves as a foundational benchmark for evaluating the predictive performance of the proposed RETO model. This dataset encompasses diverse vehicle geometries, including sedans, trucks, and SUVs. The ground-truth flow fields are generated by solving the steady state RANS equations using a finite element solver, complemented by the $k-\epsilon$ turbulence model \cite{1990The} to obtain pressure distributions. Under consistent flow conditions, the inlet velocity is fixed at $U_{\infty} = 20$ m/s, yielding a characteristic Reynolds number of $Re = 5 \times 10^6$. For the supervised learning process, the dataset is partitioned into 440 training samples, 55 validation samples, and 55 test samples, providing a robust basis for assessing the model’s generalization across varying aerodynamic profiles.

\begin{center}
    \begin{minipage}{0.6\linewidth}
        \centering
        \includegraphics[width=\linewidth]{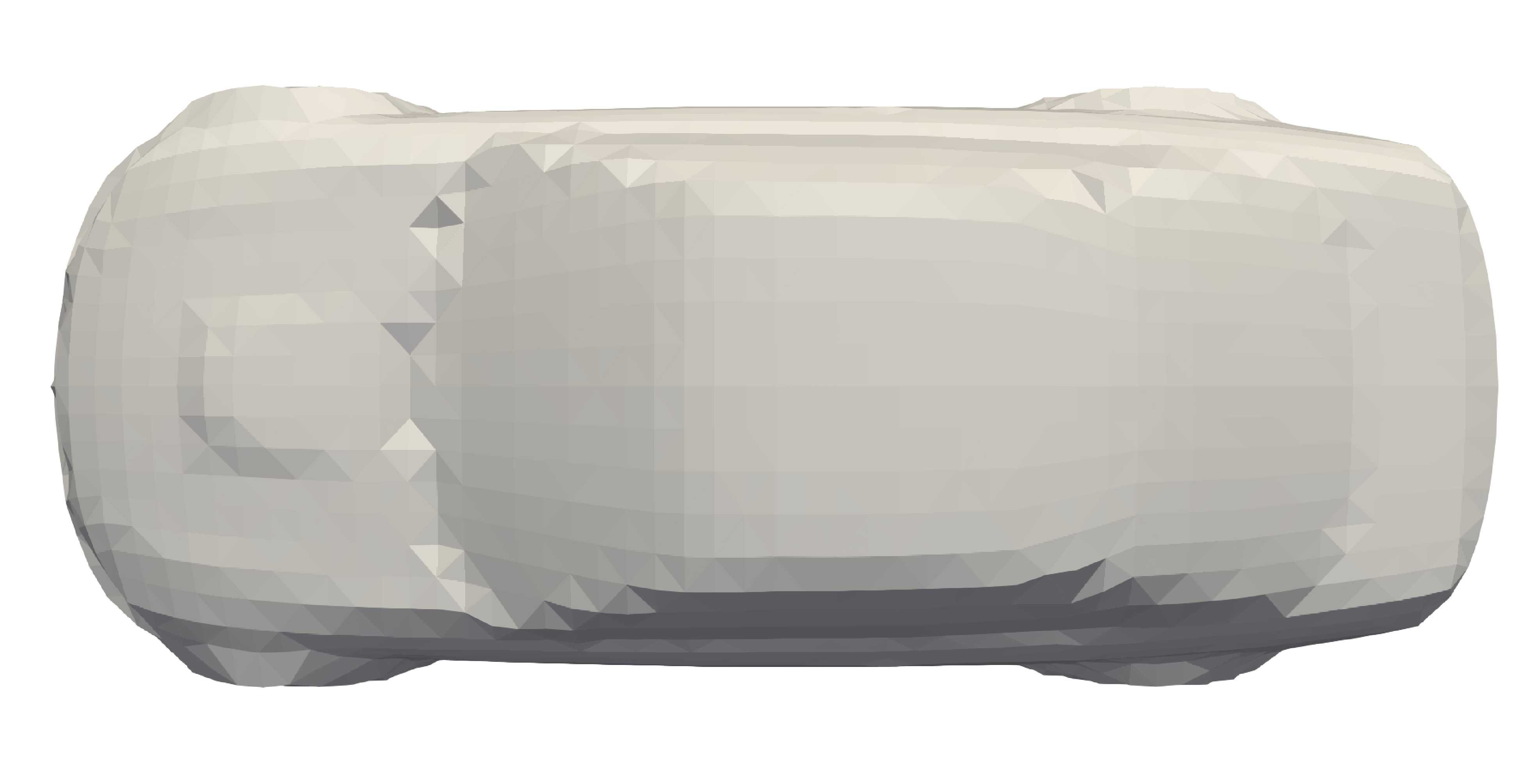}
        (a) Side view
    \end{minipage}
    \begin{minipage}{0.6\linewidth}
        \centering
        \includegraphics[width=\linewidth]{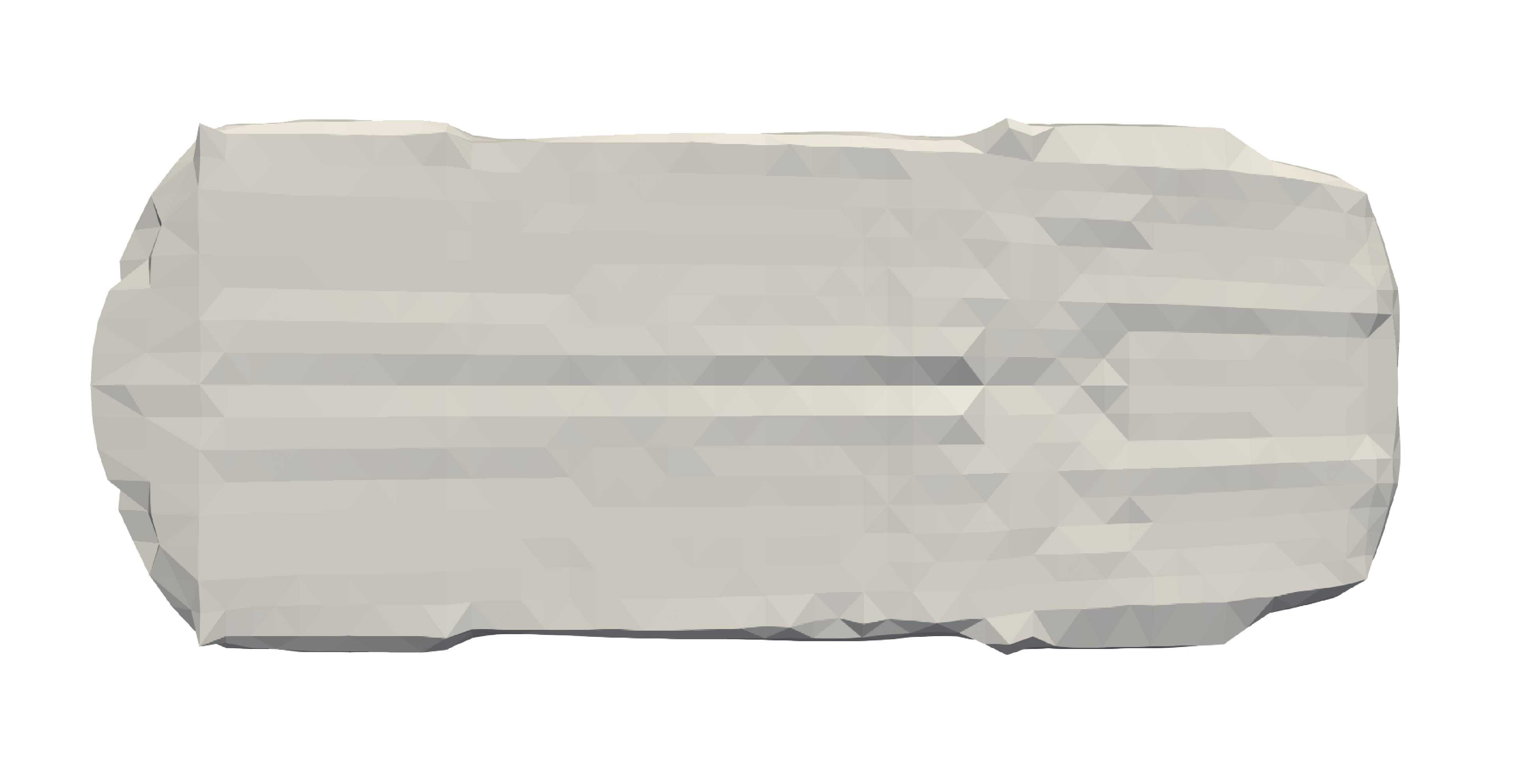}
        (b) Bottom view
    \end{minipage}
    \begin{minipage}{0.6\linewidth}
        \centering
        \includegraphics[width=\linewidth]{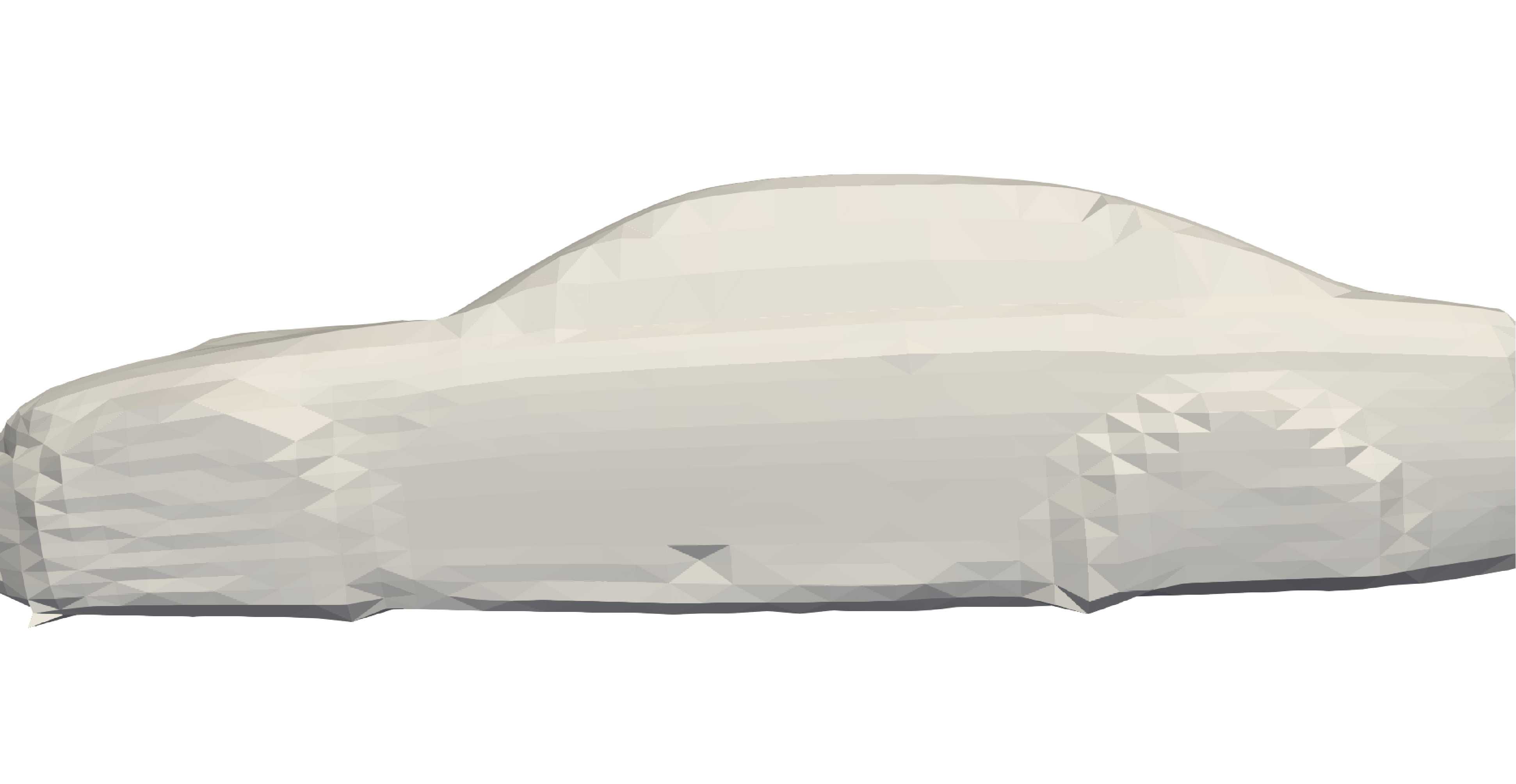}
        (c) Top view
    \end{minipage}
    \captionof{figure}{Geometric details of the ShapeNet.}
    \label{fig:ShapeNet_geo}
\end{center}

The geometric details of the ShapeNet model are illustrated in Fig.\ref{fig:ShapeNet_geo}, showing the side, bottom, and top views. Discretized by approximately $5\times10^3$ mesh nodes per manifold, ShapeNet dataset provides a lightweight benchmark for fast iteration and preliminary performance assessment. This level of geometric fidelity allows for the rapid evaluation of large-scale neural operator architectures, focusing the model learning capacity on the pressure distributions dictated by the vehicle global silhouette.

\begin{center}
    \begin{minipage}{0.7\linewidth}
        \centering
        \includegraphics[width=\linewidth]{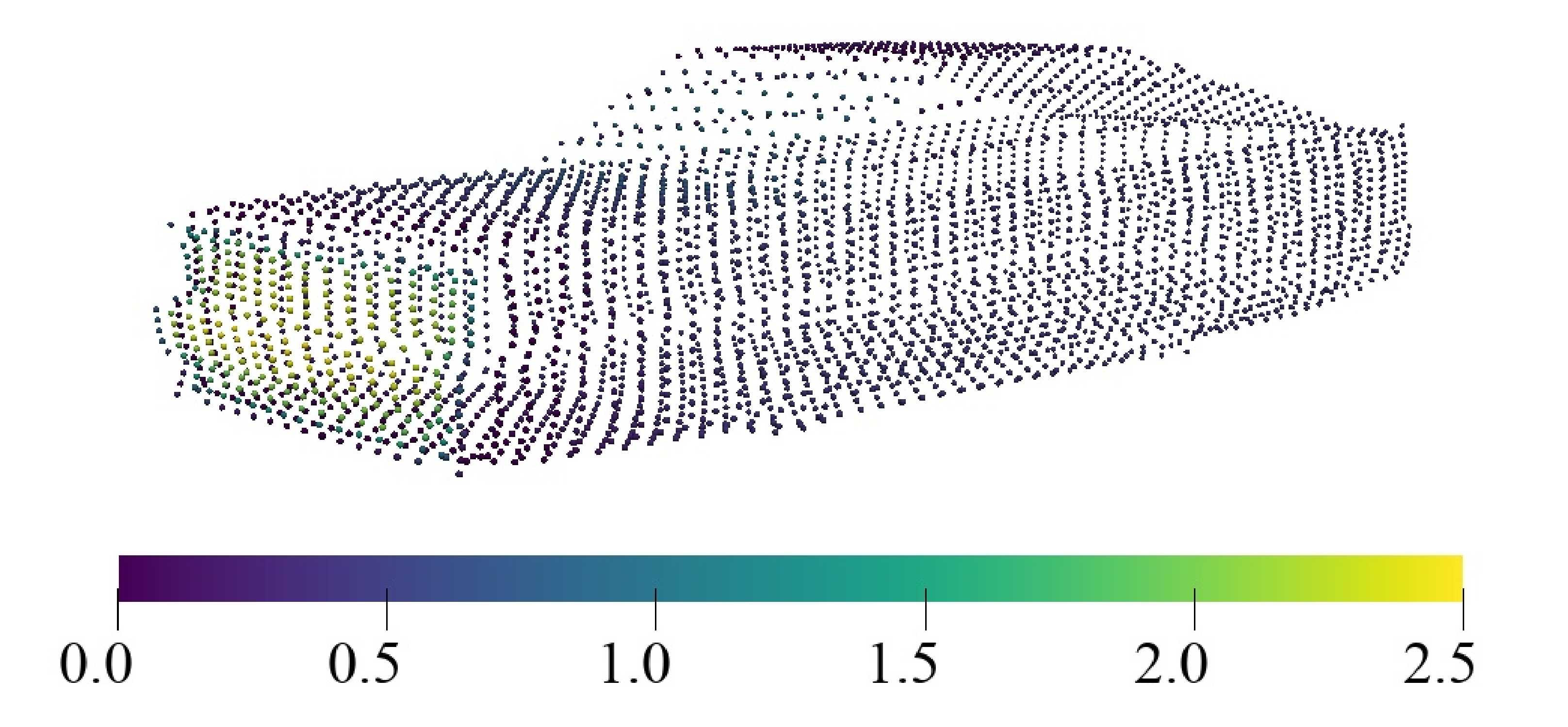}
        (a) Normalized surface pressure
    \end{minipage}
    \vspace{0.5em}
    \begin{minipage}{0.7\linewidth}
        \centering
        \includegraphics[width=\linewidth]{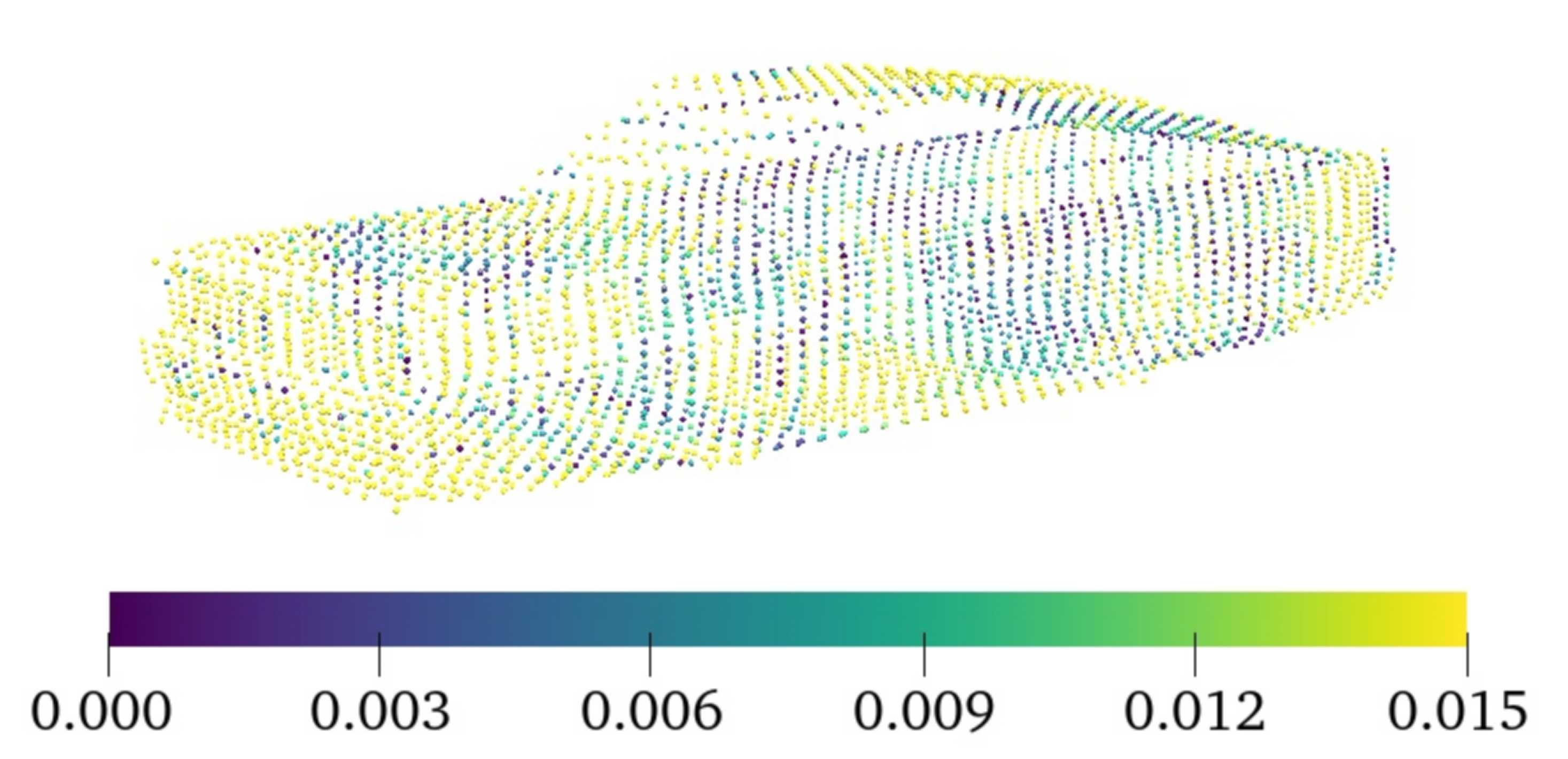}
        (b) Transolver error
    \end{minipage}
    \vspace{0.5em}
    \begin{minipage}{0.7\linewidth}
        \centering
        \includegraphics[width=\linewidth]{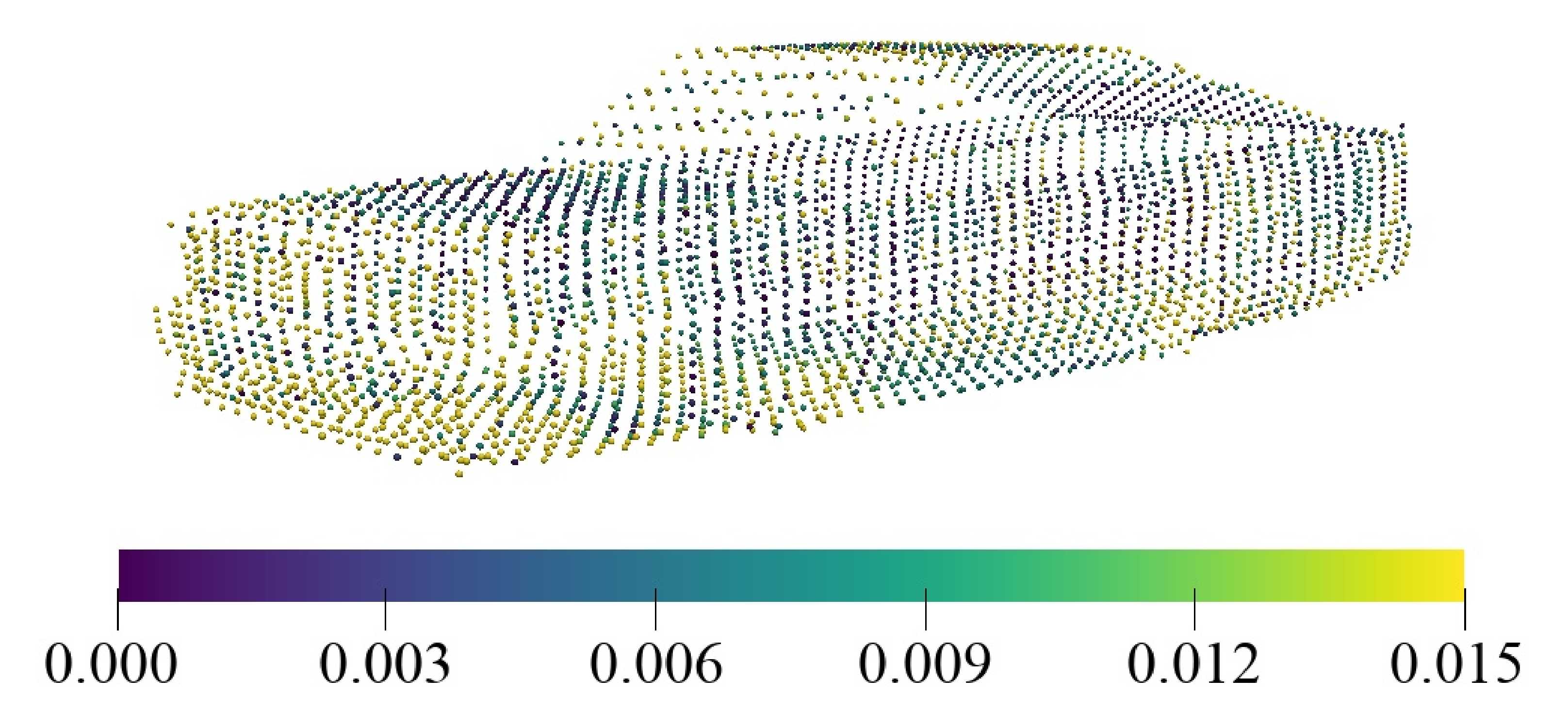}
        (c) RETO error
    \end{minipage}

    \captionof{figure}{Visualization of pressure prediction results on the ShapeNet dataset: (a) shows the ground truth field, while (b) and (c) present the reconstruction errors of Transolver and the proposed RETO, respectively.}
    \label{fig:ShapeNet_vis}
\end{center}

The comparative performance of the RETO against the baseline Transolver is quantitatively summarized in Table \ref{tab:model_ShapeNet} and qualitatively illustrated in Fig.\ref{fig:ShapeNet_vis}. The proposed RETO demonstrates superior predictive performance on the ShapeNet dataset compared to existing models. While RegDGCNN and Transolver yield relative $L_2$ errors of 0.125 and 0.075 respectively, RETO reduces the surface pressure error to 0.063. This 16\% enhancement in predictive fidelity over Transolver demonstrates the superior capacity of the rotary-enhanced attention in mapping complex geometric coordinates to physical pressure states.

\begin{table*}
\centering
\caption{Relative L2 error across models on ShapeNet}
\begin{tabular*}{\textwidth}{@{\extracolsep{\fill}} l c c c}
\hline
Model       & Pressure($L_2$)  \\
\hline
RegDGCNN    & 0.125     \\
Transolver  & 0.075     \\
RETO        & 0.063     \\
\hline
\end{tabular*}
\label{tab:model_ShapeNet}
\end{table*}

\begin{center}
  \includegraphics[width=0.5\textwidth]{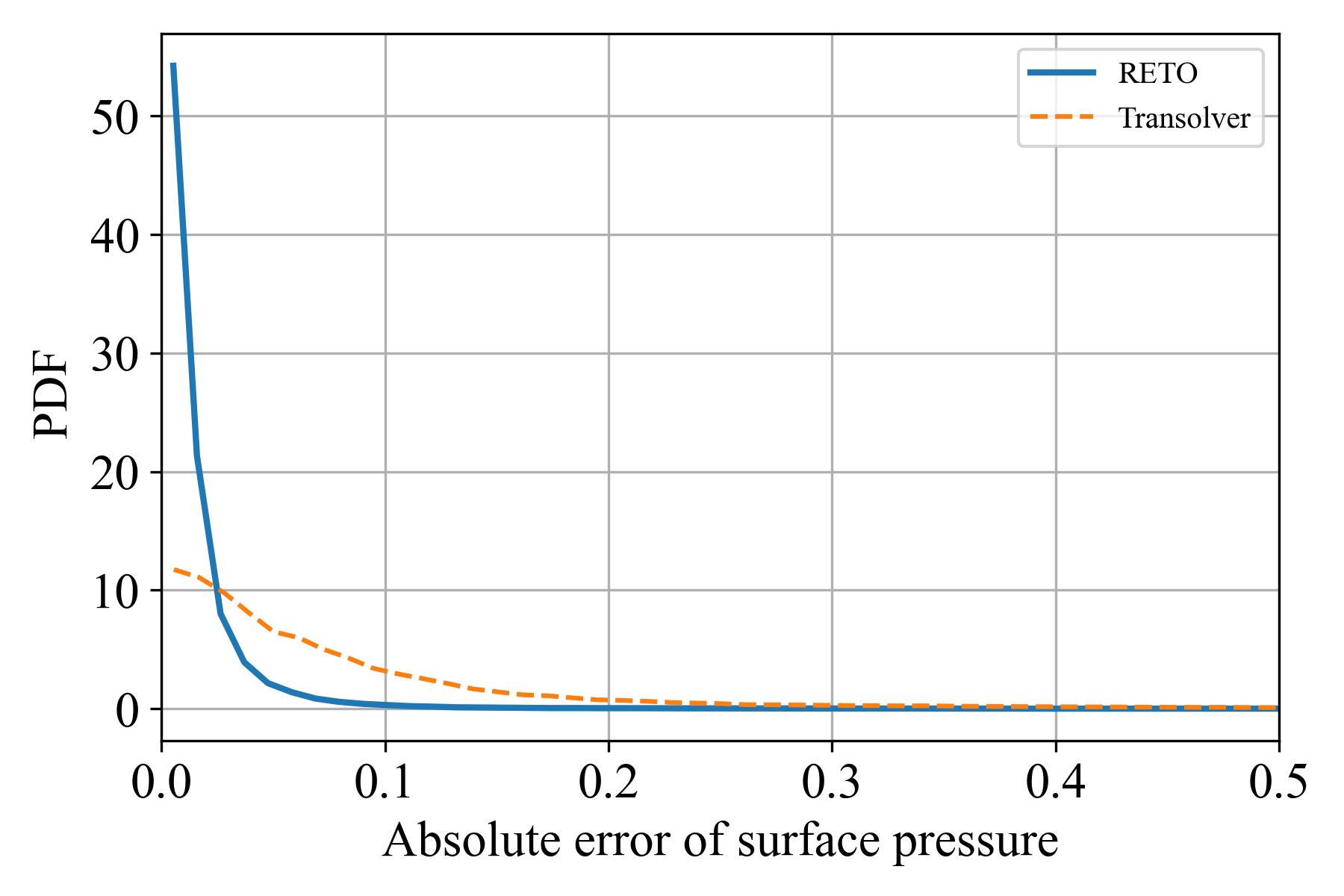}
  \captionof{figure}{PDF of error of surface pressure predictions for the ShapeNet dataset.}
  \label{fig:ShapeNet_PDFcurve}
\end{center}

To further scrutinize the predictive performance, the probability density functions (PDFs) of the absolute errors for the surface pressure, aggregated over all samples in the ShapeNet test set, are compared in Fig.\ref{fig:ShapeNet_PDFcurve}. Note that all error metrics are computed using normalized dimensionless fields to ensure numerical consistency across the dataset. A quantitative analysis of the PDF profiles reveals several critical advantages of the proposed RETO model. The error distribution of RETO is characterized by a remarkably sharp and high-magnitude peak near the origin, with the probability density reaching approximately 55. In contrast, the Transolver baseline reaches a peak density of only approximately 12 in the same near-zero regime. This substantial disparity in peak intensity indicates that RETO achieves near-ground-truth accuracy for a significantly larger proportion of the spatial nodes across the entire test suite. Furthermore, the overall error mass of RETO is more tightly clustered toward zero, exhibiting a rapid decay as the error magnitude increases. While Transolver displays a much broader distribution with a non-negligible tail of errors spanning the $[0.1, 0.25]$ range, RETO effectively suppresses these moderate-magnitude errors. This high concentration of probability density suggests that the proposed model maintains higher localized precision and superior statistical robustness across the complex vehicle manifold, ensuring high-fidelity reconstruction even in regions with sensitive physical gradients.

\begin{center}
\begin{minipage}{0.8\linewidth}
    \centering
    \includegraphics[width=\linewidth]{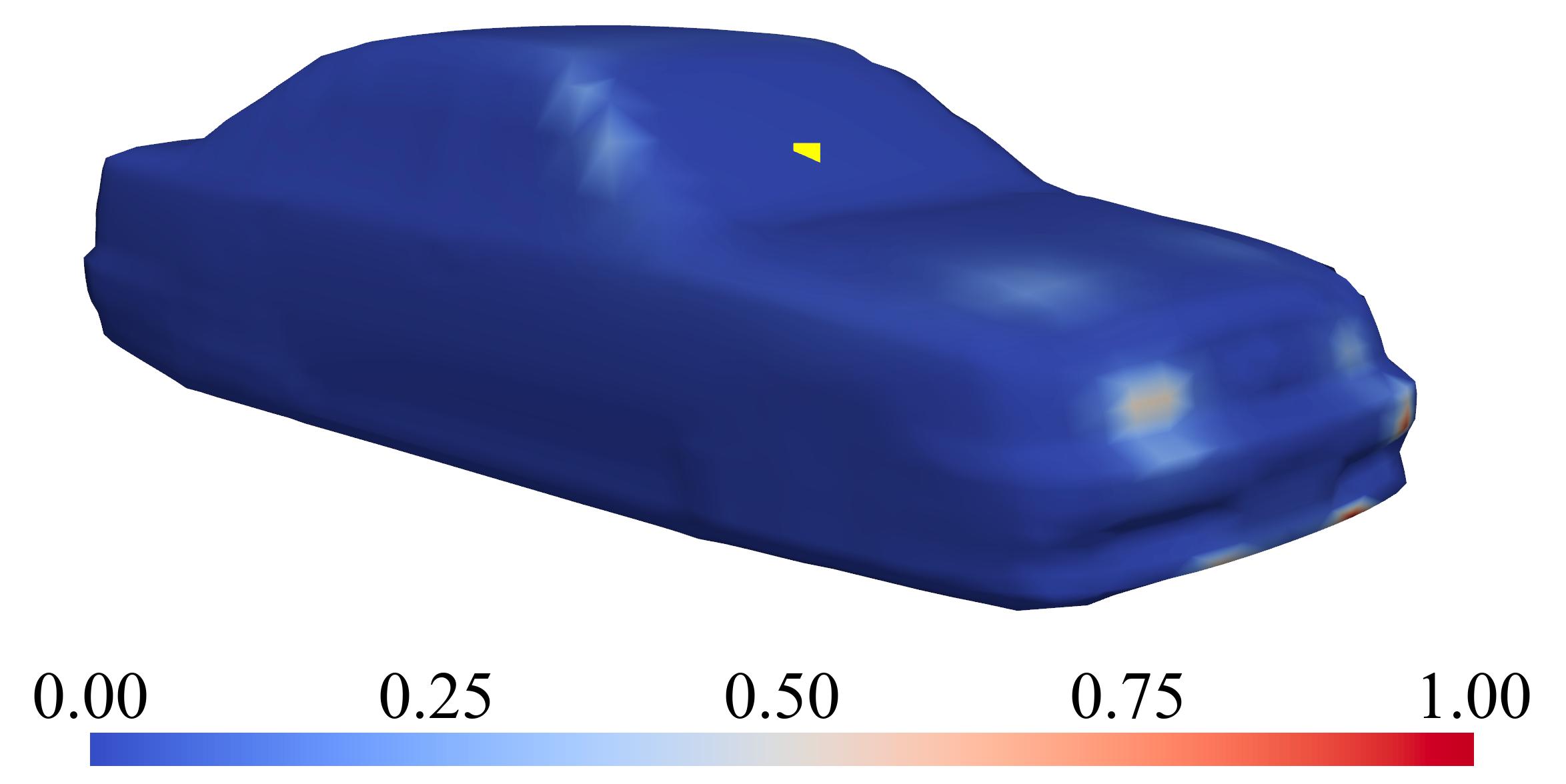}
    (a) Global view
\end{minipage}
\begin{minipage}{0.8\linewidth}
    \centering
    \includegraphics[width=\linewidth]{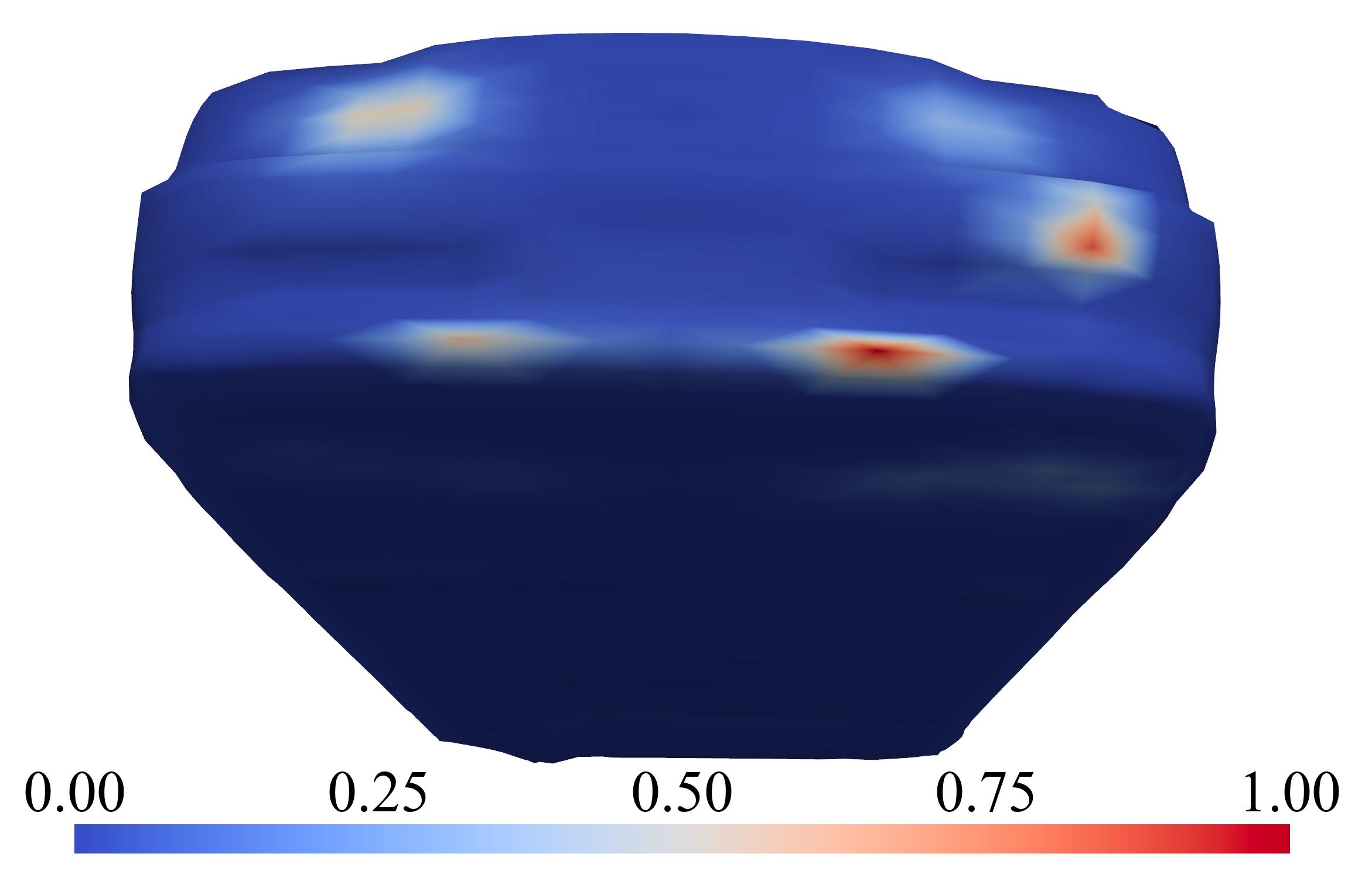}
    (b) Zoom-in view
\end{minipage}
\captionof{figure}{Visualization of the 3D Spatial Attention Kernel in RETO.}
\label{fig:ShapeNet_attentionMap}

\end{center}

Fig.\ref{fig:ShapeNet_attentionMap} provides a qualitative visualization of the 3D attention kernel \cite{tsai2019transformer} within the RETO framework. By selecting a query point on the front windshield (indicated by the yellow marker), the attention distribution across the vehicle manifold is mapped. The results reveal a spatially coherent and localized attention field. High attention weights are not only concentrated in the immediate vicinity of the query point but also propagate along the A-pillars and the front hood. This demonstrates that RETO successfully captures the geometric continuity of the vehicle, recognizing that the aerodynamic performance of the windshield is physically coupled with the upstream flow from the hood and the lateral flow separation at the A-pillars.

\begin{figure*}
    \centering
    \includegraphics[width=0.8\textwidth]{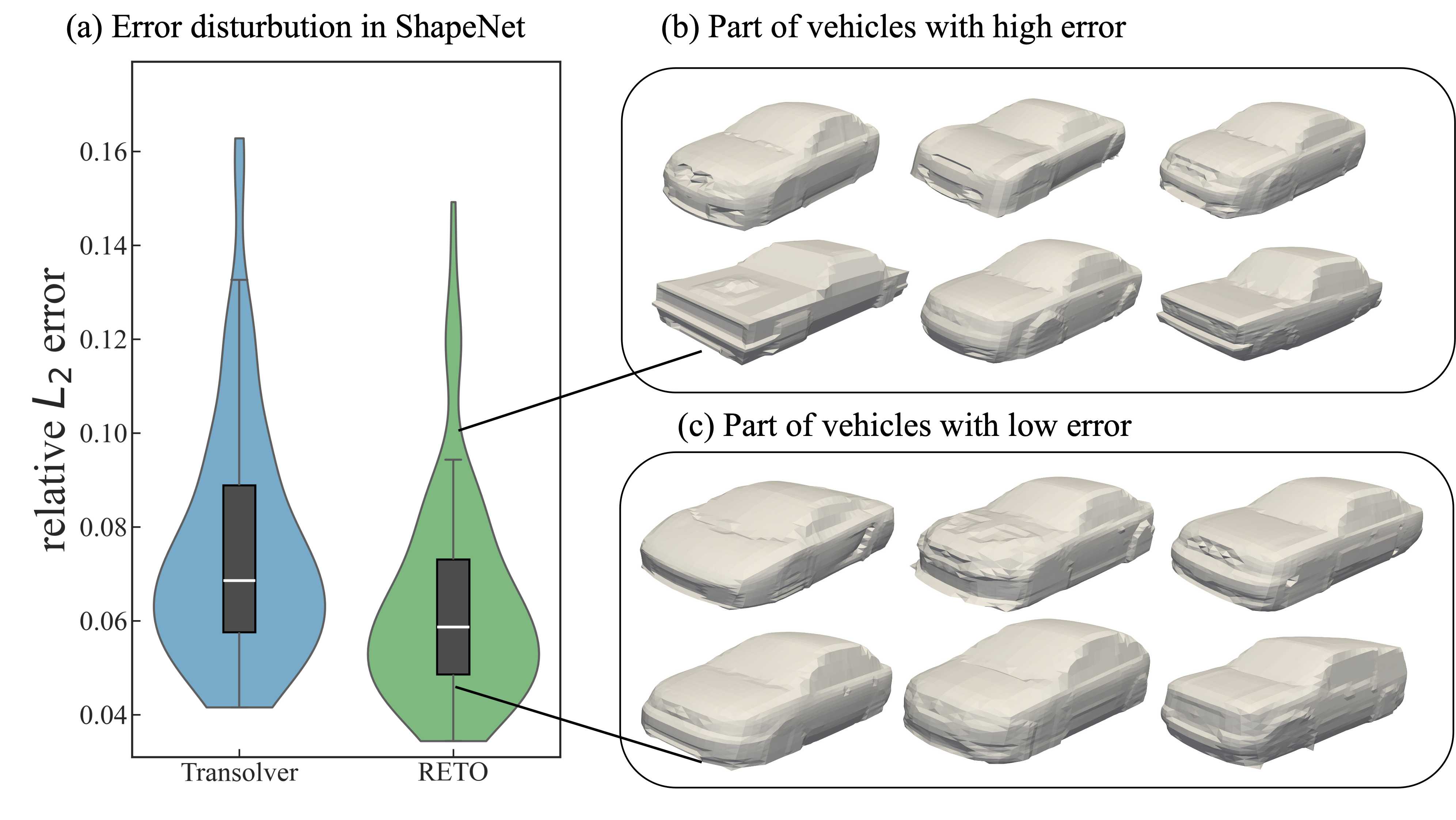}
    \caption{(a) The surface pressure error distribution of Transolver and RETO for $L_2$ error in ShapeNet Car. 
    The six vehicle types with the highest (b) and lowest (c) errors in the test set using the RETO model.}
    \label{fig:violin_ShapeNet}
\end{figure*}

In Fig.\ref{fig:violin_ShapeNet} (a), the predictive performance of the RETO model is evaluated against the Transolver baseline using the relative $L_2$ error distribution on the ShapeNet test set. The results demonstrate that RETO significantly outperforms the baseline across all statistical quartiles. Notably, the more pronounced reduction in the upper quartile and the overall error dispersion suggests that the proposed architecture is more robust in handling geometrically challenging samples and mitigating extreme prediction errors. To provide qualitative context for these statistical observations, Figures \ref{fig:violin_ShapeNet} (b) and (c) illustrate representative vehicle geometries from the high and low-error regimes, respectively. The high-error instances shown in (b) are primarily associated with unconventional vehicle topologies, such as boxy vintage designs characterized by sharp geometric discontinuities and irregular proportions. Such geometries pose substantial challenges for spatial feature extraction due to the complex flow separation and intense pressure gradients they induce. Conversely, the low-error cases in (c) predominantly feature streamlined aerodynamic profiles typical of modern sedans and sports cars. The consistently lower error on these geometries indicates that RETO effectively captures the fundamental physical patterns of conventional vehicle aerodynamics. This improvement reflects that the integration of RoPE and optimized attention mechanisms enhances the model’s ability to generalize across the diverse ShapeNet manifold, specifically by mitigating predictive failures on non-standard and complex geometries.

\subsection{DrivAerML}
The DrivAerML dataset \citep{ashton2024drivaerml} represents the most physically demanding benchmark in this study, characterized by its high-resolution representation of intricate flow structures. Based on the DrivAer notchback with the open-cooling (OCDA) configuration \cite{hupertz2021aerodynamics}, the dataset comprises 500 parametric variations across 16 design parameters, including A-pillar angles and cooling inlet geometries. The ground-truth simulations solve the incompressible Navier-Stokes equations using OpenFOAM (v2212), employing DDES \cite{spalart2009detached} with a spalart-allmaras background model. Unlike steady-state RANS, the DDES approach resolves large-scale turbulent eddies within the wake, providing a significantly more accurate yet complex training target for the neural operator. Under a fixed inlet velocity of $U_{\infty} = 38.9$ m/s, the flow reaches a characteristic Reynolds number of $Re_L = 7.19 \times 10^6$, with each sample computational mesh consisting of approximately 160 million cells. For the experimental setup, the data is partitioned into 400 training, 50 validation, and 50 test samples to rigorously evaluate RETO predictive performance in industrial-grade aerodynamic scenarios.

\begin{figure*}
\centering
\subfloat[Side view]{%
  \includegraphics[width=0.32\linewidth]{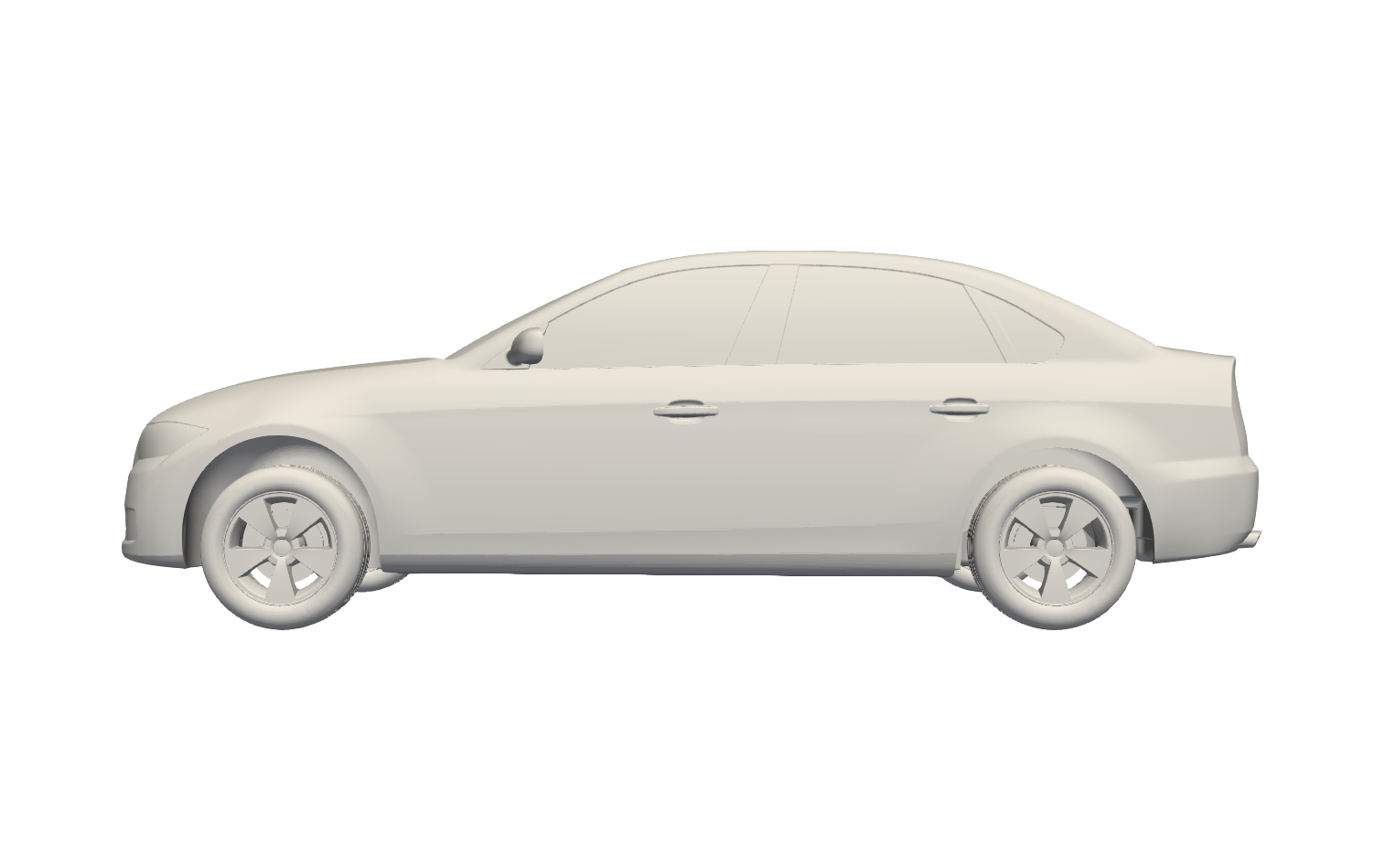}%
  \label{fig:v3}%
}
\hfill
\subfloat[Bottom view]{%
  \includegraphics[width=0.32\linewidth]{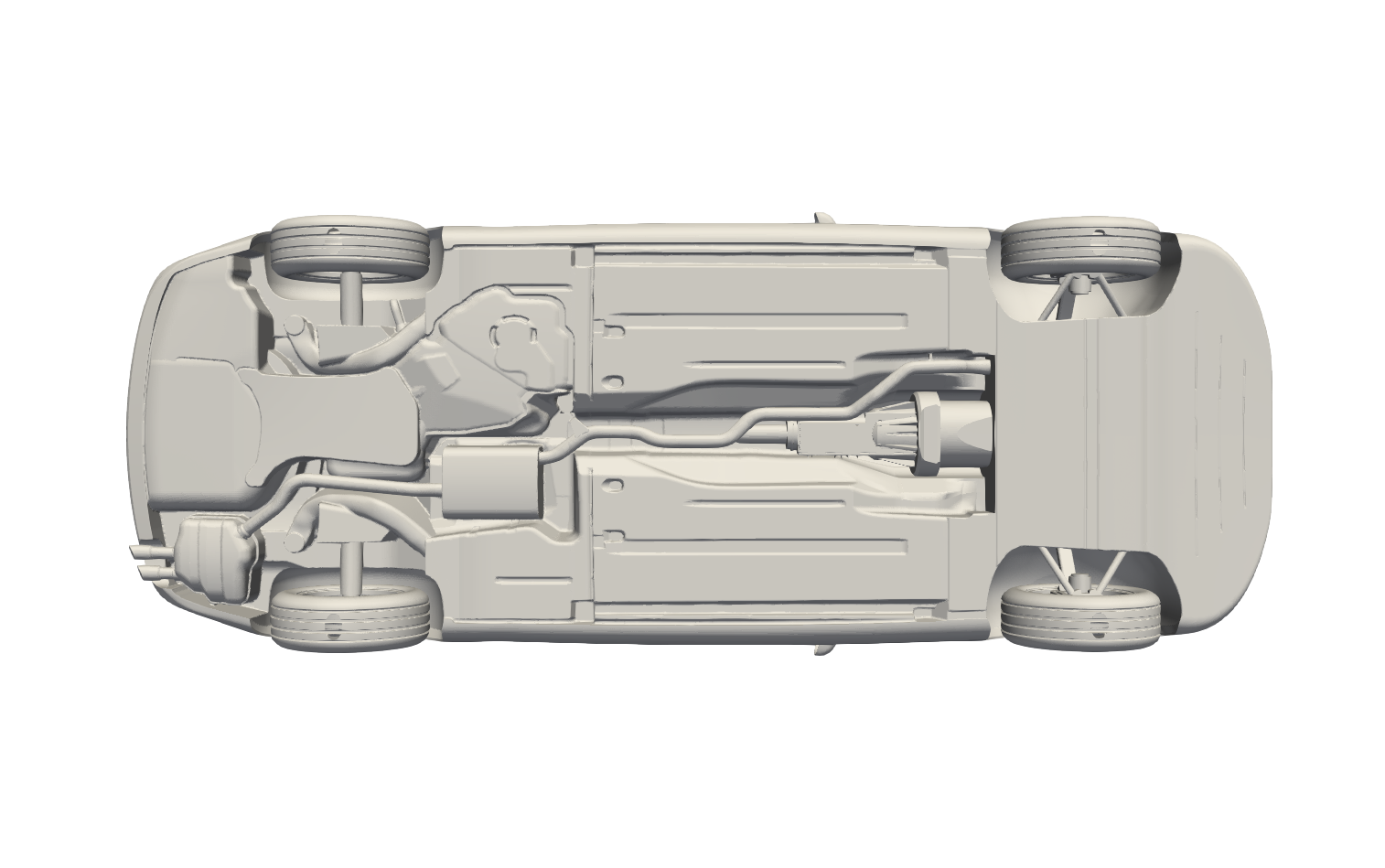}%
  \label{fig:geom_side_v1}%
}
\hfill
\subfloat[Top view]{%
  \includegraphics[width=0.32\linewidth]{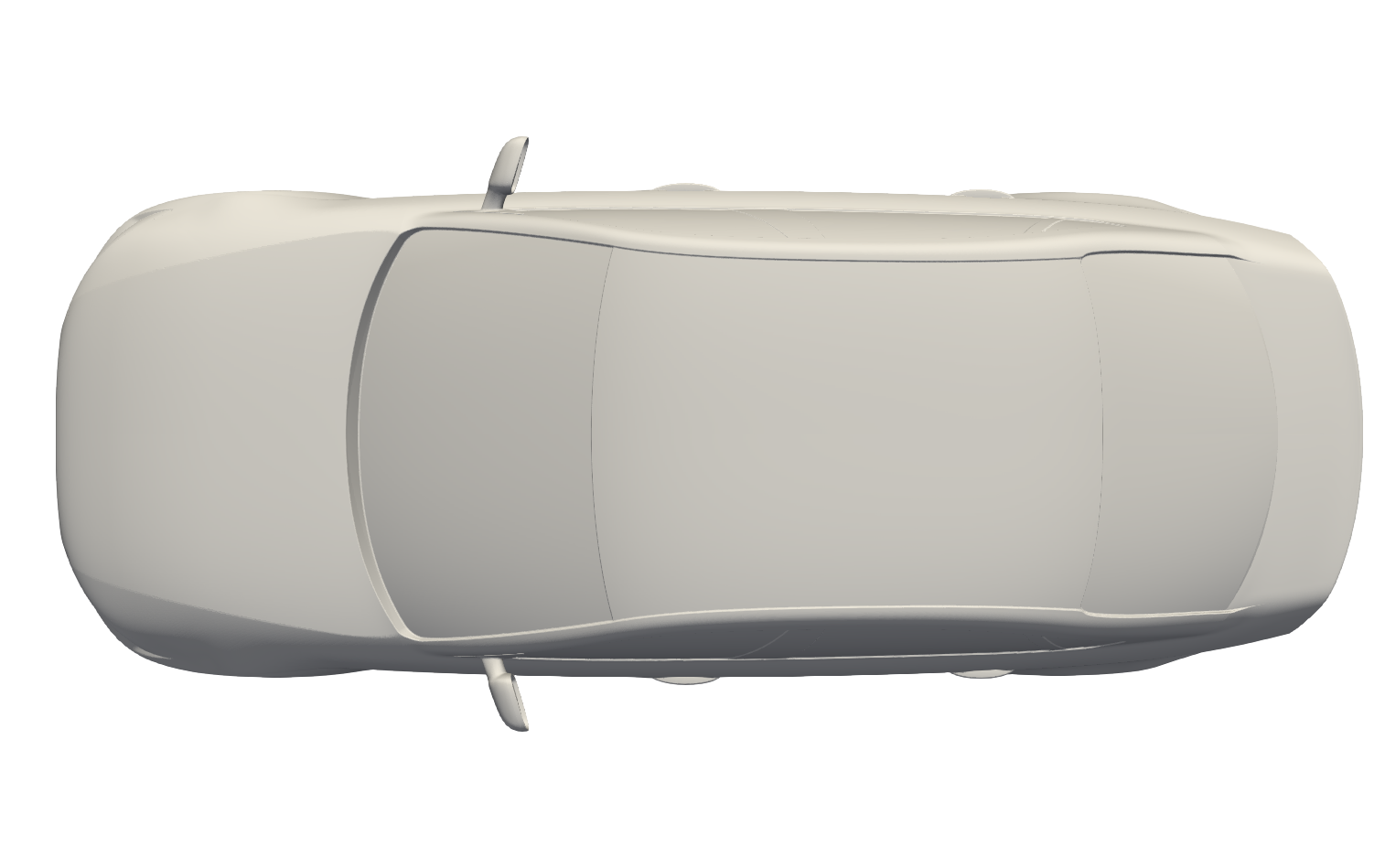}%
  \label{fig:geom_detail_v1}%
}
\caption{Geometric details of the vehicle model.}
\label{fig:vehicle_geometry}
\end{figure*}

\begin{figure*}
\centering
\subfloat[Pressure]{%
  \includegraphics[width=0.48\linewidth]{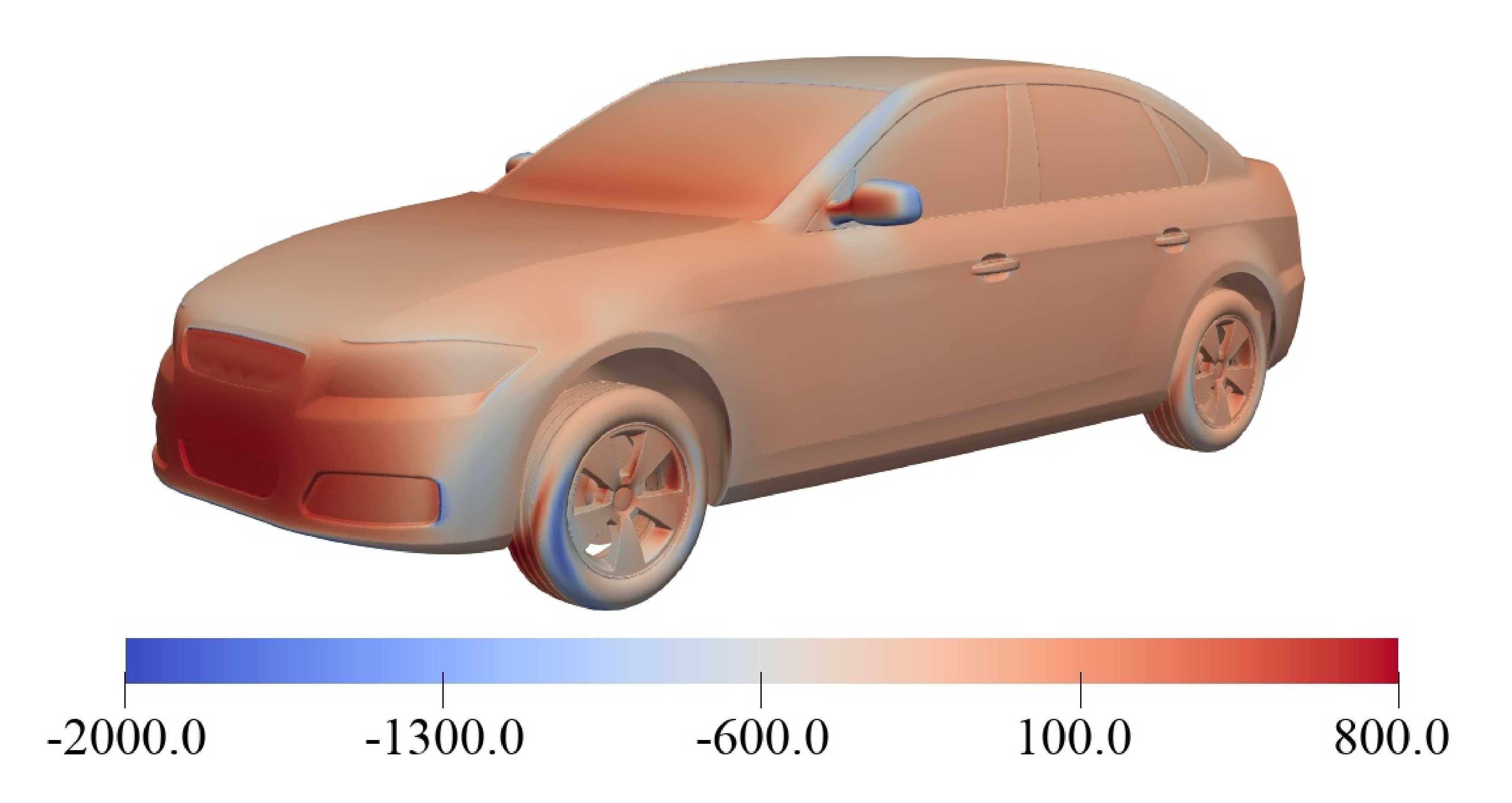}%
  \label{fig:pressure}%
}
\hfill
\subfloat[Streamlines coloured by velocity magnitude]{%
  \raisebox{0pt}{
    \includegraphics[width=0.48\linewidth]{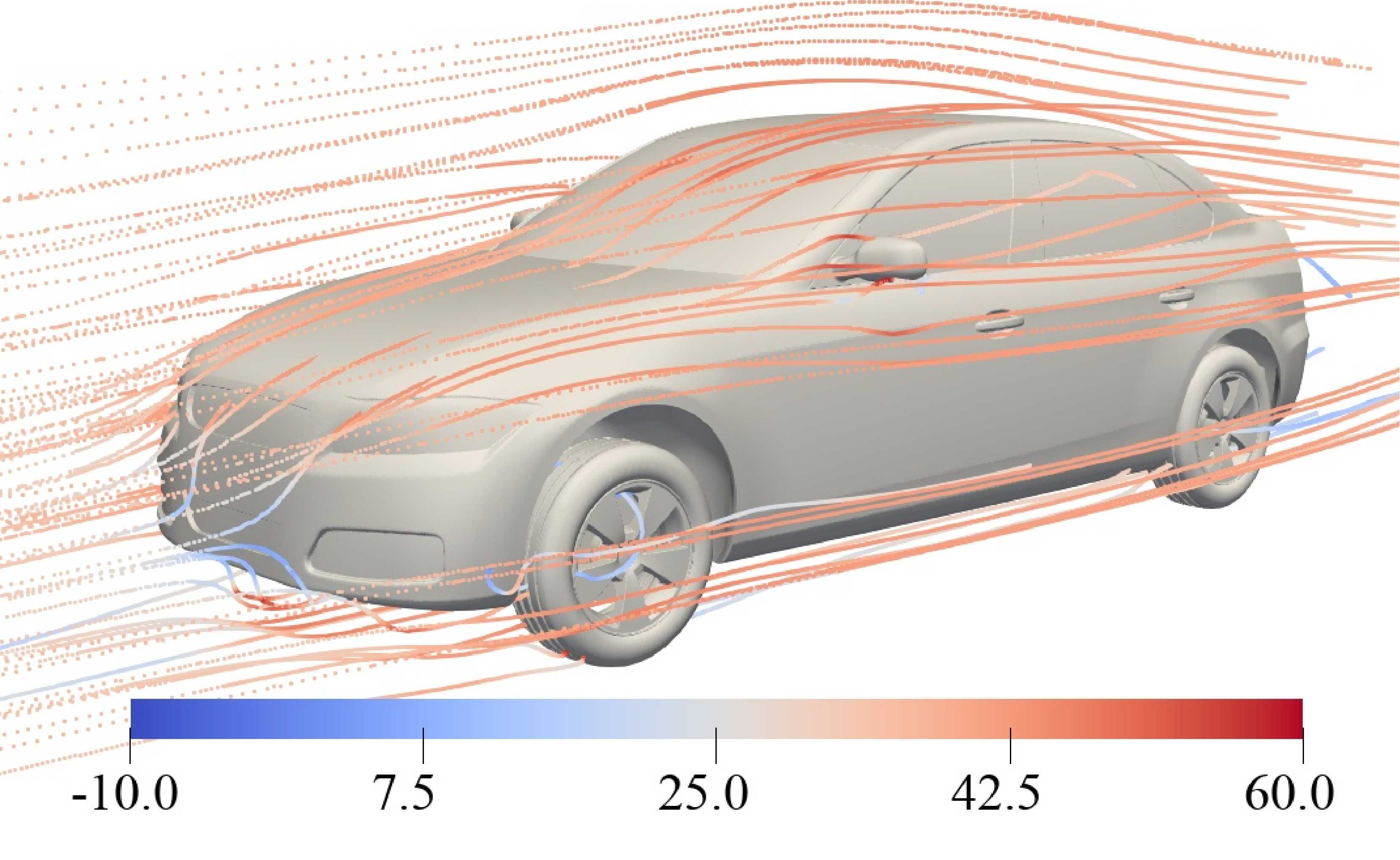}
  }%
  \label{fig:velocity}%
}
\caption{Flow field visualization of the vehicle model.}
\label{fig:flow_fields}
\end{figure*}

\begin{table*}
\centering
\caption{Relative $L_2$ error and ablation study on the DrivAerML dataset using $10^4$ random points}
\label{tab:DrivAerML_result}
\begin{tabular*}{\textwidth}{@{\extracolsep{\fill}} l c c}
\hline
Model  & Pressure ($L_2$) & Velocity ($L_2$) \\
\hline
RegDGCNN      & 0.235          & 0.312 \\ 
AB-UBT        & 0.102          & 0.124 \\
Transolver    & 0.116          & 0.121 \\
Transolver++  & 0.127           & 0.133 \\
Transolver-3  & 0.121           & 0.136 \\
\textbf{RETO} & \textbf{0.089} & \textbf{0.097} \\
\hline
\textit{Ablation Study:} & & \\
v1: w/o RoPE & 0.104 & 0.101 \\
v2: w/o sin-cos & 0.660 & 0.680 \\
v3: w/o RoPE \& sin-cos & 0.200 & 0.190 \\
\hline
\end{tabular*}
\end{table*}

\begin{table*}
\centering
\caption{Efficiency comparison for surface pressure prediction on DrivAerML}
\begin{tabular*}{\textwidth}{@{\extracolsep{\fill}} l c c c}
\hline
Model       & Inference Latency (s) & Parameters (M) & Memory (GB) \\
\hline
RegDGCNN    & 0.02                   & 0.14          & 1.24        \\
Transolver  & 0.02                   & 0.17          & 0.08         \\
Transolver++  & 0.06                   & 0.21          & 0.12         \\
Transolver-3  & 0.05                   & 0.20          & 0.10         \\
AB-UBT      & 2.89                   & 1.32          & 0.82         \\
RETO        & 0.09                   & 0.33          & 0.10   \\
\hline
\end{tabular*}
\label{tab:efficiency_drivAerML}
\end{table*}

The geometric intricacies of the DrivAerML model are illustrated in Fig.~\ref{fig:vehicle_geometry}, showing the side, bottom, and top views. These details impose significant challenges for the neural operator, as the model must capture how subtle geometric variations modulate the downstream aerodynamic performance. Fig.\ref{fig:flow_fields} (a) depicts the predicted surface pressure distribution, highlighting the high-pressure stagnation zone at the front and localized low-pressure regions near aerodynamic features such as the side mirrors and wheel arches. The streamlines in Fig.\ref{fig:flow_fields} (b) reveal the intricate flow patterns around the DrivAerML vehicle. As the air passes over the streamlined body, the lines remain predominantly attached, yet they converge and accelerate at the frontal pillars and roof curvature. 

The quantitative performance of the proposed RETO model on the DrivAerML test set is summarized in Table \ref{tab:DrivAerML_result}. The relative $L_2$ errors for surface pressure and velocity are reported as $0.235$/$0.312$ for RegDGCNN~\cite{2024DrivAerNet}, $0.102$/$0.124$ for AB-UBT~\cite{alkin2025ab}. For Transolver~\cite{wu2024transolver}, Transolver++~\cite{luo2025transolver++}, and Transolver-3~\cite{zhou2026transolver}, the respective errors for surface pressure and velocity are $0.116$/$0.121$, $0.127$/$0.133$, and $0.121$/$0.136$. This trend is consistent with the recent survey CarBench~\cite{elrefaie2025carbench}, which also reports that Transolver++ underperforms Transolver in terms of prediction accuracy. In comparison, RETO gives the highest accuracy with $0.089$/$0.097$. The results demonstrate that the full RETO architecture achieves superior predictive fidelity, outperforming the baseline Transolver by 29\% in surface pressure and 19\% in volumetric velocity fields. The ablation study confirms that the dual-stage encoding is the primary driver of accuracy. The ablation study in Table \ref{tab:DrivAerML_result} further dissects the contribution of each component. Specifically, for surface pressure prediction, RETO achieves a relative $L_2$ error that is 14\% lower than v1, 86\% lower than v2, and 55\% lower than v3. This indicates that RoPE indeed plays a positive role in the overall architecture. The anomalously high error of v2 should be attributed to the instability introduced by the removal of other components, rather than the ineffectiveness of RoPE itself. Overall, the combination of RoPE and sin-cosine positional encodings employed in RETO proves to be effective and well-suited for the aerodynamic field reconstruction task.

Table \ref{tab:efficiency_drivAerML} compares inference efficiency and model size on DrivAerML surface pressure prediction. All results are measured on a single NVIDIA Tesla V100-SXM2-32GB GPU with batch size 1. While RETO is not the fastest among the compared models, it achieves the highest prediction accuracy, demonstrating a favorable trade-off between computational cost and fidelity. This makes RETO well-suited for applications where accuracy is prioritized over raw inference speed.

\begin{figure*}
    \centering
    \subfloat[Normalized surface pressure]{
        \includegraphics[width=0.3\linewidth]{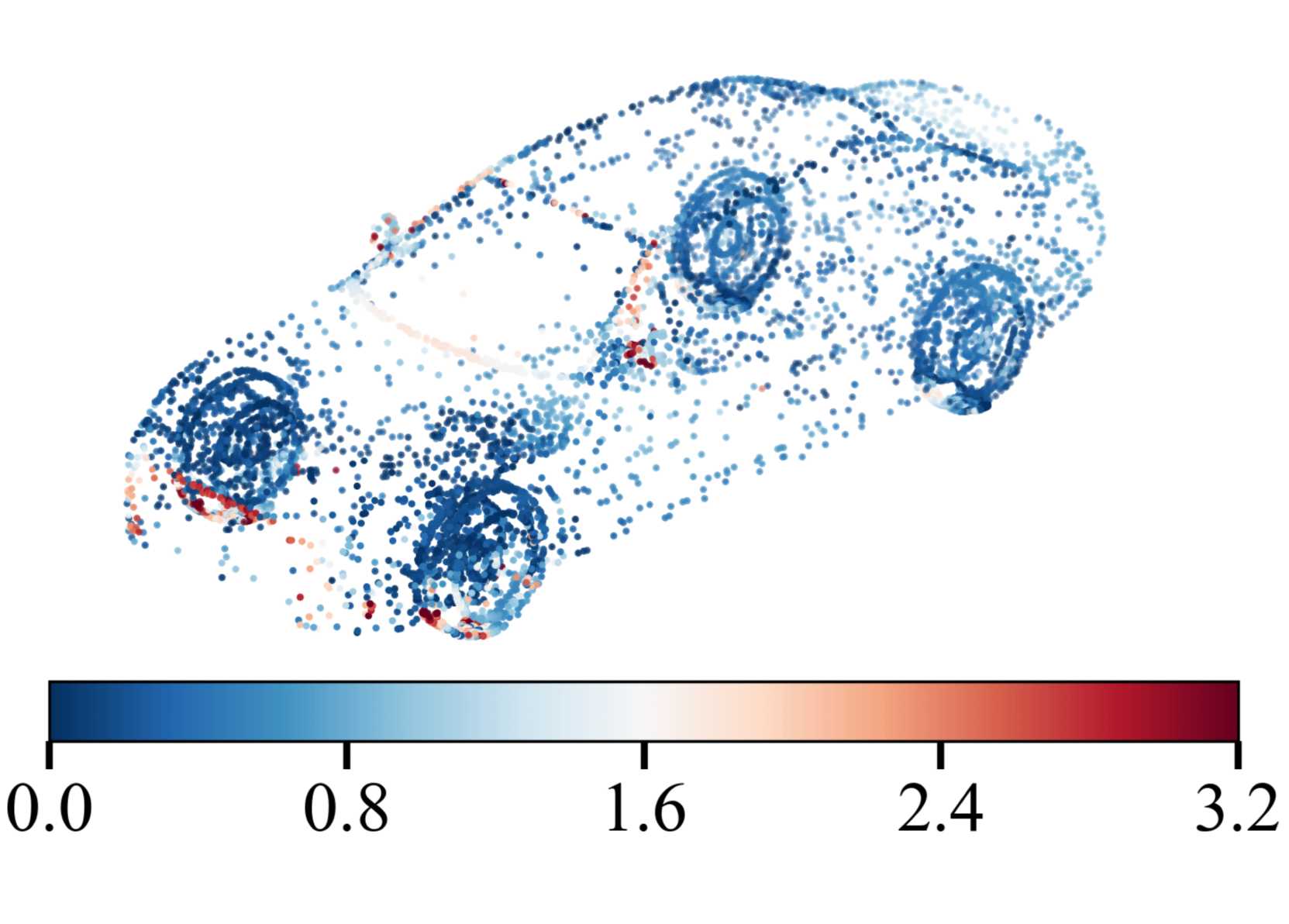}
        \label{fig:pressure_true}
    }  
    \subfloat[Transolver error]{
        \includegraphics[width=0.3\linewidth]{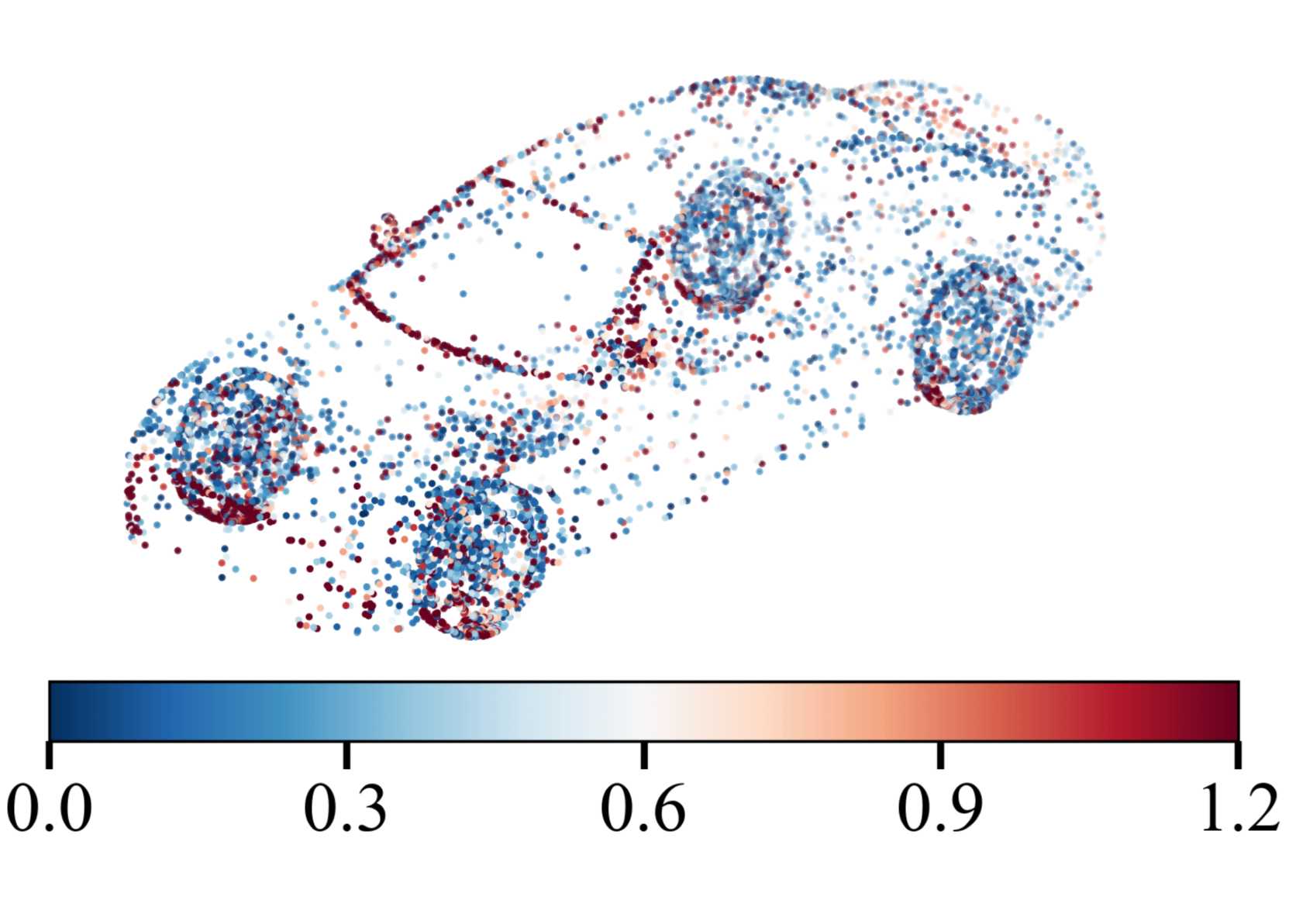}
        \label{fig:pressure_transolver_error}
    }
    \subfloat[RETO error]{
        \includegraphics[width=0.3\linewidth]{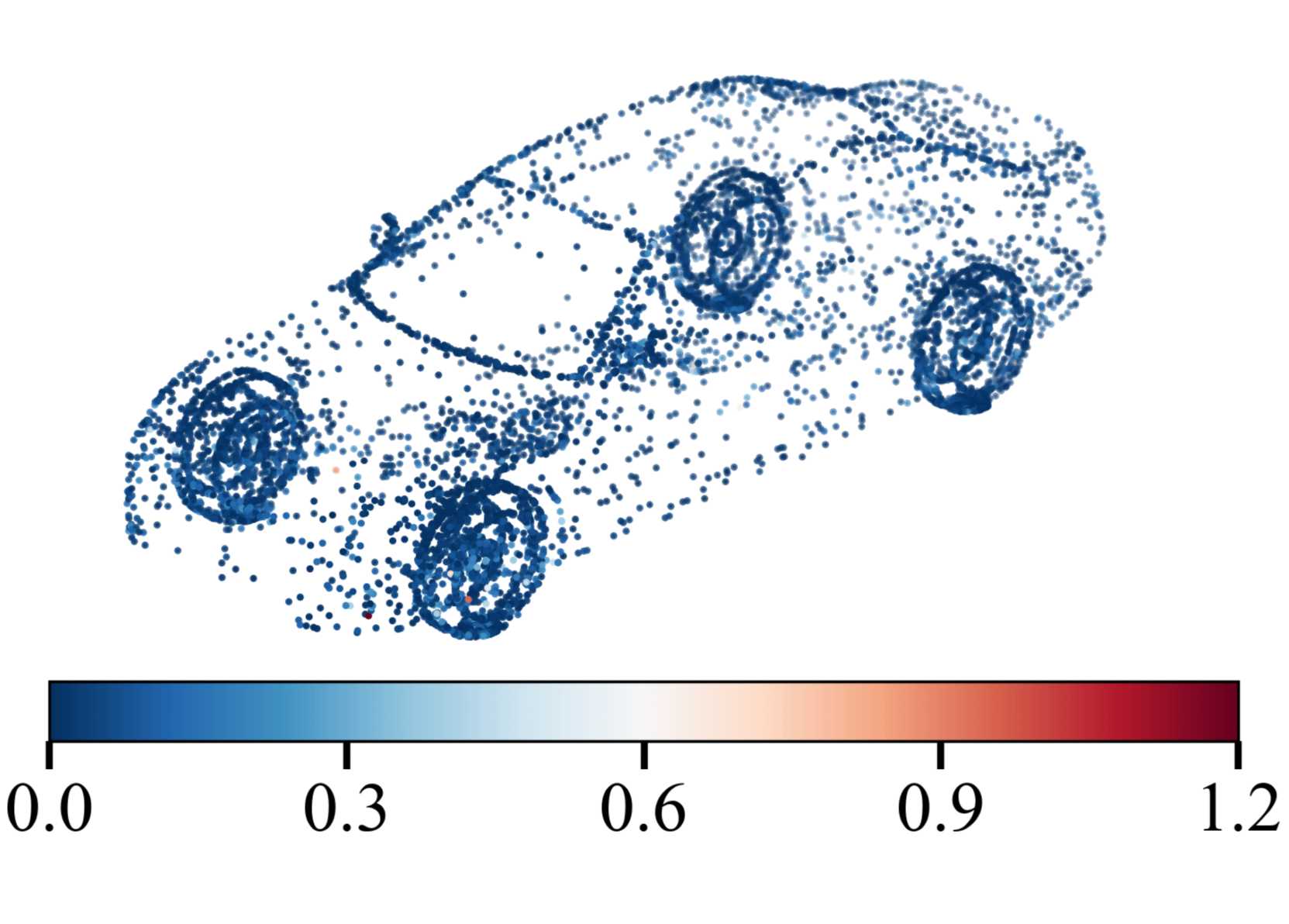}
        \label{fig:pressure_RETO_error}
    }
    
    \caption{Visualization of predicted surface pressure fields and absolute errors sampled at $10^4$ random points for (a) Normalized surface pressure, (b) Transolver, and (c) RETO.}
    \label{fig:DrivAerML_pressure_vis}
\end{figure*}

\begin{figure*}
    \centering
    \includegraphics[width=0.8\textwidth]{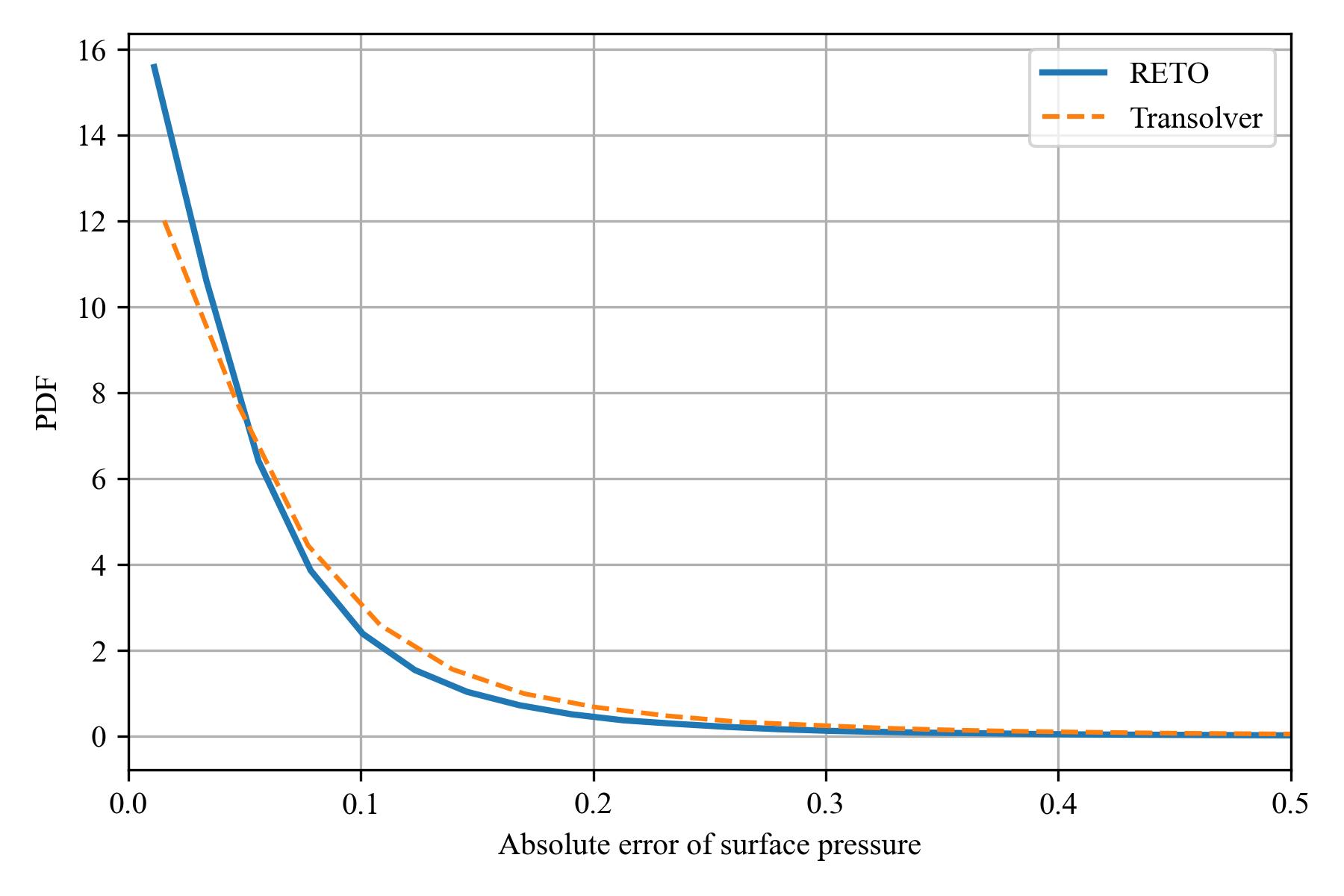}
    \caption{PDF of error of surface pressure predictions for the DrivAerML dataset.}
    \label{fig:DrivAerML_pressure_PDFcurve}
\end{figure*}

\begin{figure*}
    \centering
    \subfloat[Normalized  velocity magnitude]{
        \includegraphics[width=0.33\linewidth]{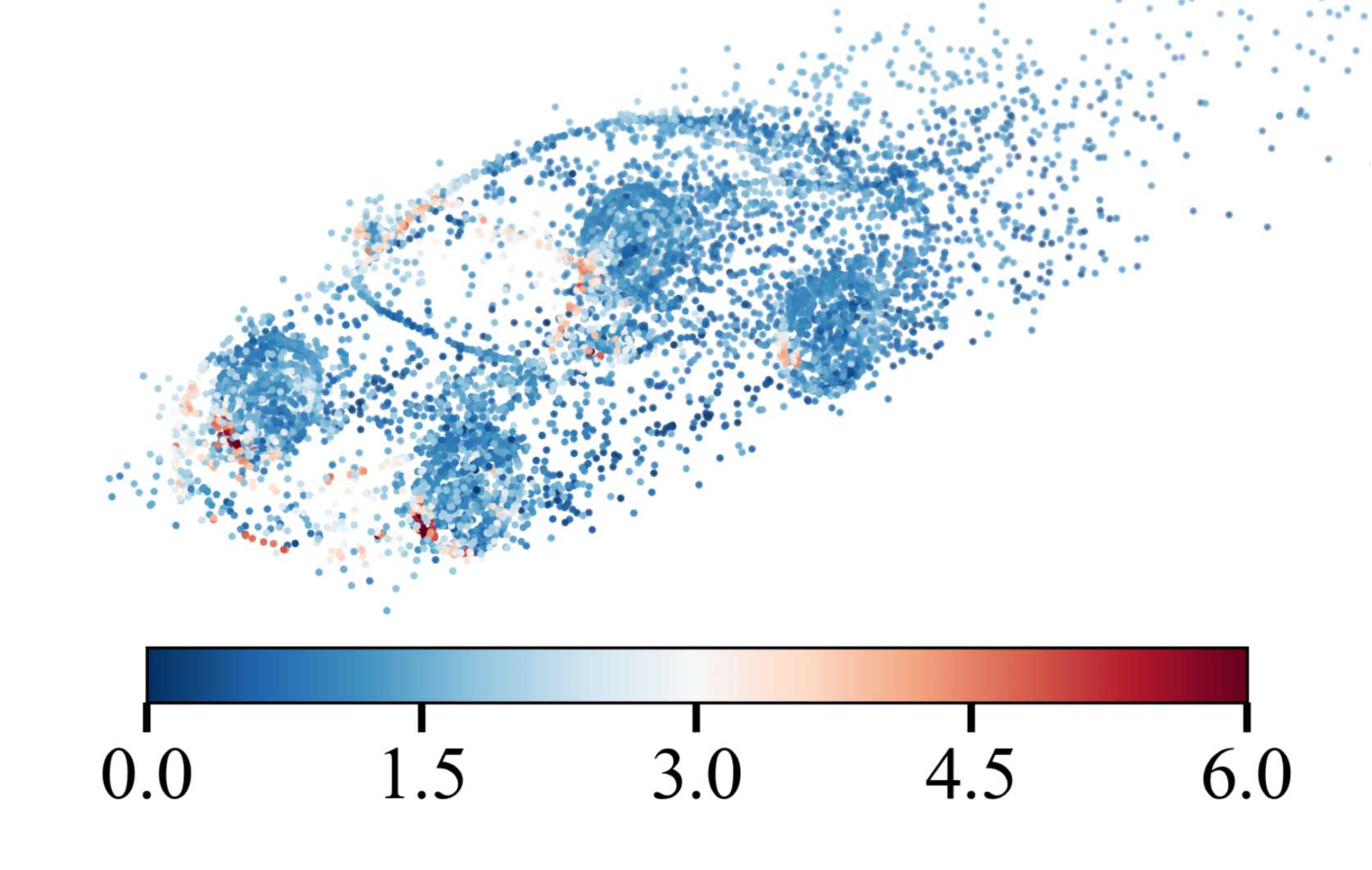}
        \label{fig:velocity_true}
    }  
    \subfloat[Transolver error]{
        \includegraphics[width=0.33\linewidth]{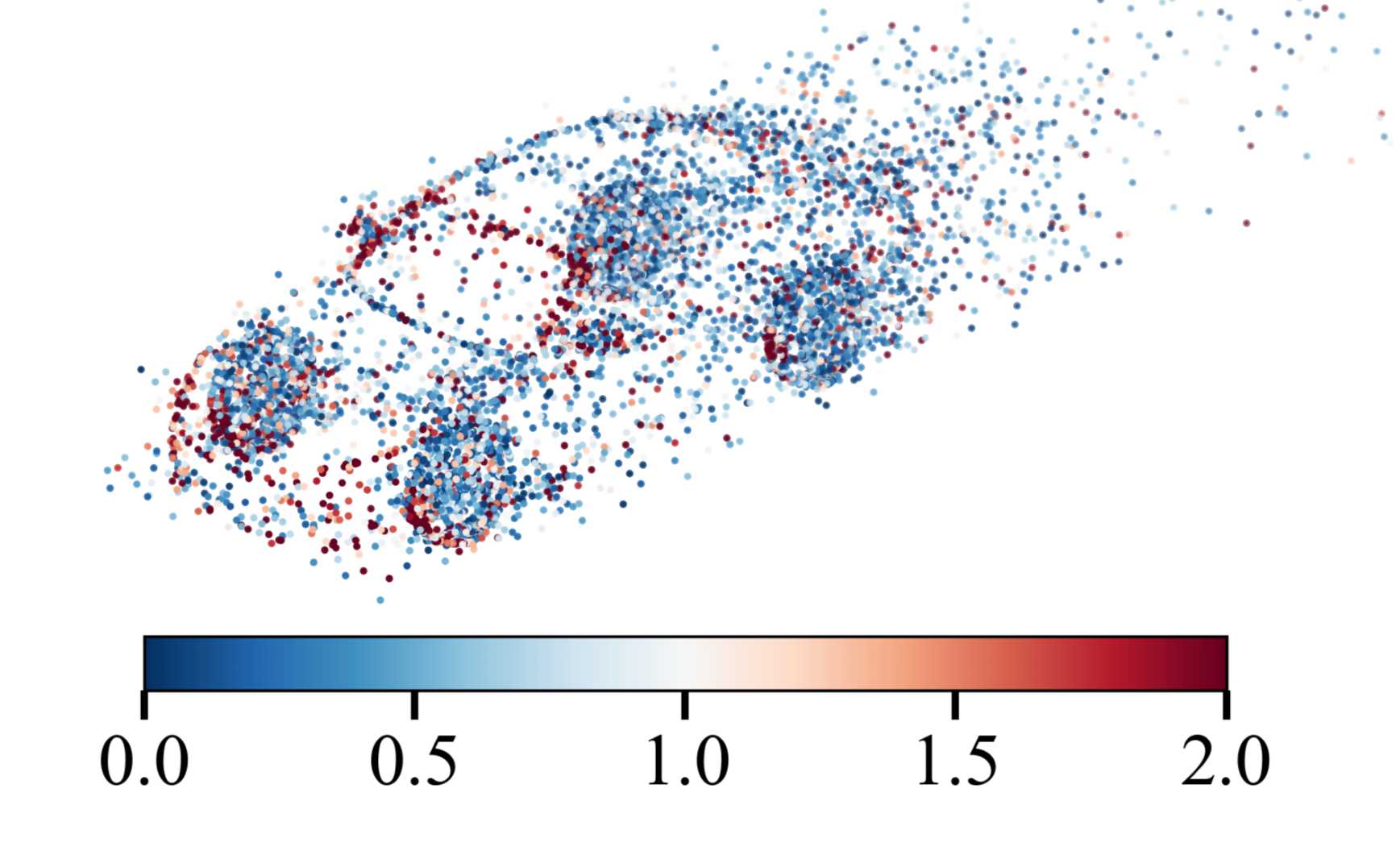}
        \label{fig:velocity_transolver_error}
    }
    \subfloat[RETO error]{
        \includegraphics[width=0.33\linewidth]{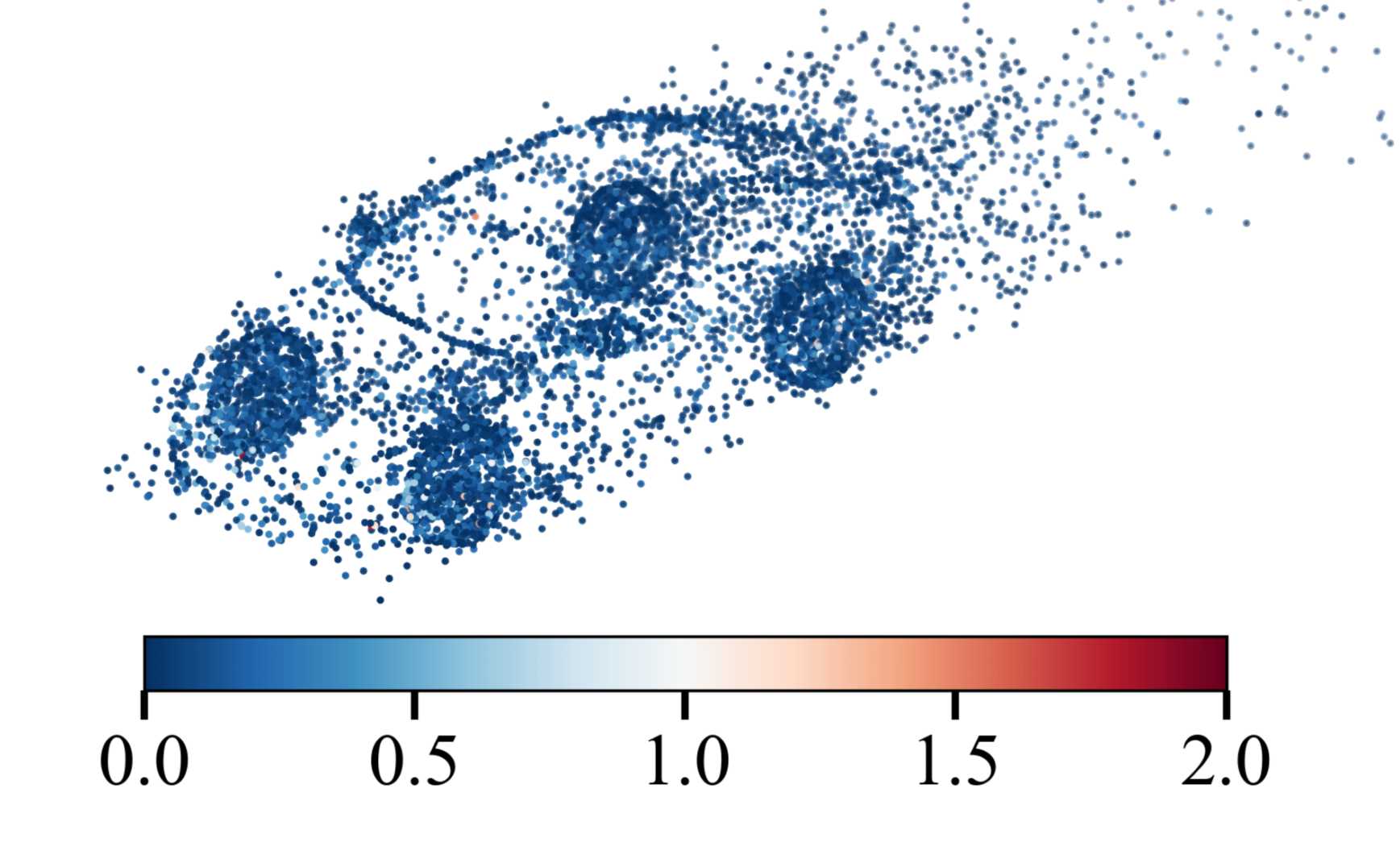}
        \label{fig:velocity_RETO_error}
    }
    
    \caption{Visualization of predicted velocity fields and absolute errors sampled at $10^4$ random points for (a) Normalized  velocity magnitude, (b) Transolver, and (c) RETO.}
    \label{fig:DrivAerML_velocity_vis}
\end{figure*}

\begin{figure*}
    \centering
    \includegraphics[width=0.8\textwidth]{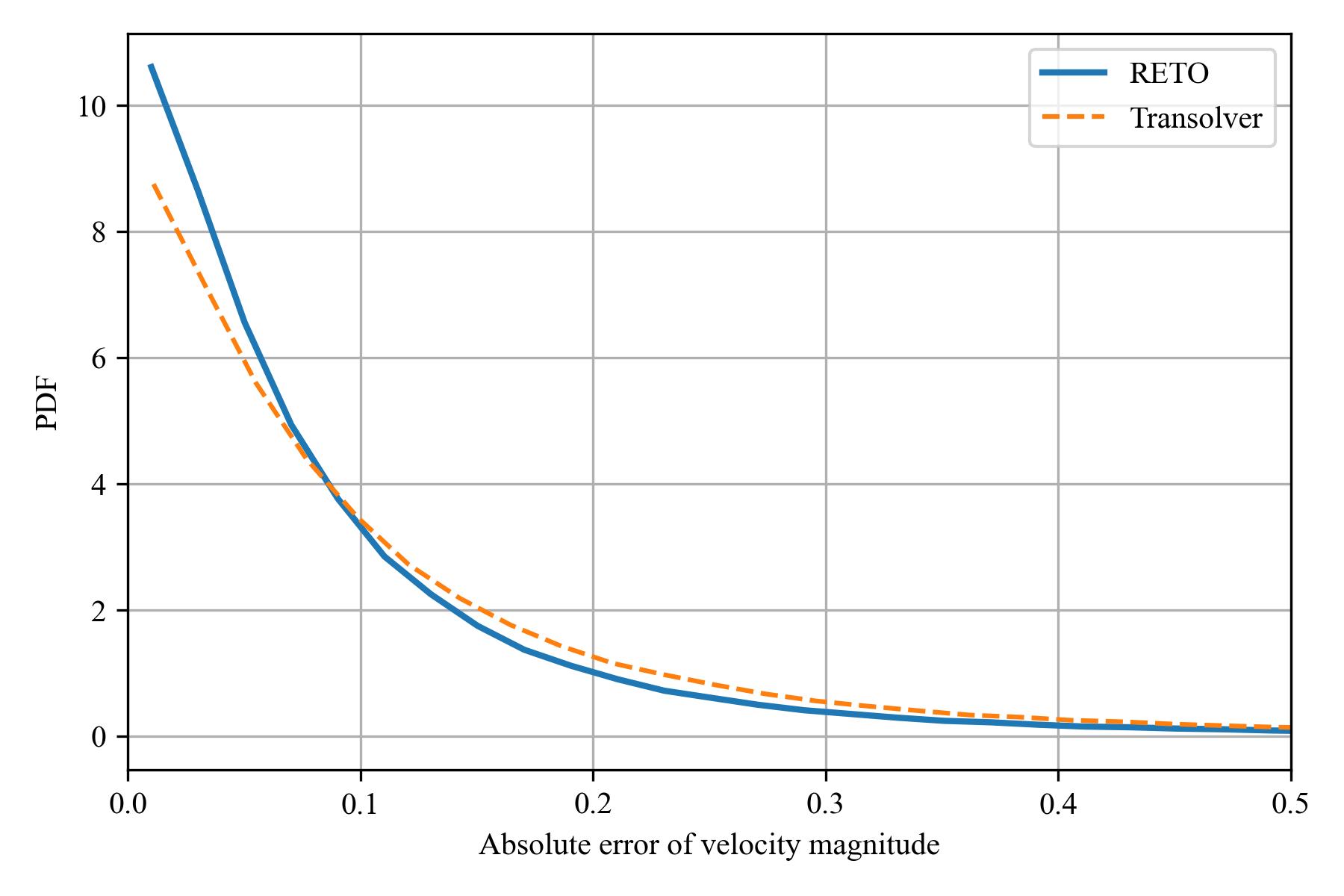}
    \caption{PDF of error of velocity magnitude predictions for the DrivAerML dataset.}
    \label{fig:DrivAerML_velocity_PDFcurve}
\end{figure*}

Qualitative visualizations in Fig.\ref{fig:DrivAerML_pressure_vis} and Fig.\ref{fig:DrivAerML_velocity_vis} further corroborate these numerical improvements. RETO effectively reconstructs sharp pressure gradients at geometric singularities—including side mirrors and A-pillars—and accurately captures the complex vortex shedding and velocity deficits within the vehicle wake. By maintaining awareness of relative spatial displacements, RETO preserves flow structures that are typically smoothed out by traditional solvers, establishing itself as a robust and high-precision surrogate for industrial aerodynamic analysis.

To further investigate the error characteristics on the DrivAerML dataset, the PDFs of absolute prediction errors for surface pressure and velocity magnitude are presented in Fig.\ref{fig:DrivAerML_pressure_PDFcurve} and Fig.\ref{fig:DrivAerML_velocity_PDFcurve}, respectively. A clear distinction in statistical performance can be observed between the two models. The error distributions of RETO exhibit a strong concentration in the low-error regime, characterized by a remarkably sharp and high-magnitude peak near the origin. Specifically, the peak density for RETO reaches approximately 15.5 for pressure and 10.5 for velocity, significantly outstripping the Transolver baseline, which manifests lower peak densities of roughly 12 and 8.8, respectively. This disparity indicates that a much larger proportion of RETO predictions achieve near-ground-truth accuracy across the complex vehicle geometry. Overall, these results demonstrate that RETO achieves superior precision in the low-error regime, and this statistical behavior suggests that the incorporation of RoPE contributes to a more consistent modeling of complex spatial dependencies in three-dimensional flow fields.
\begin{figure*}
    \centering
    \includegraphics[width=0.7\textwidth]{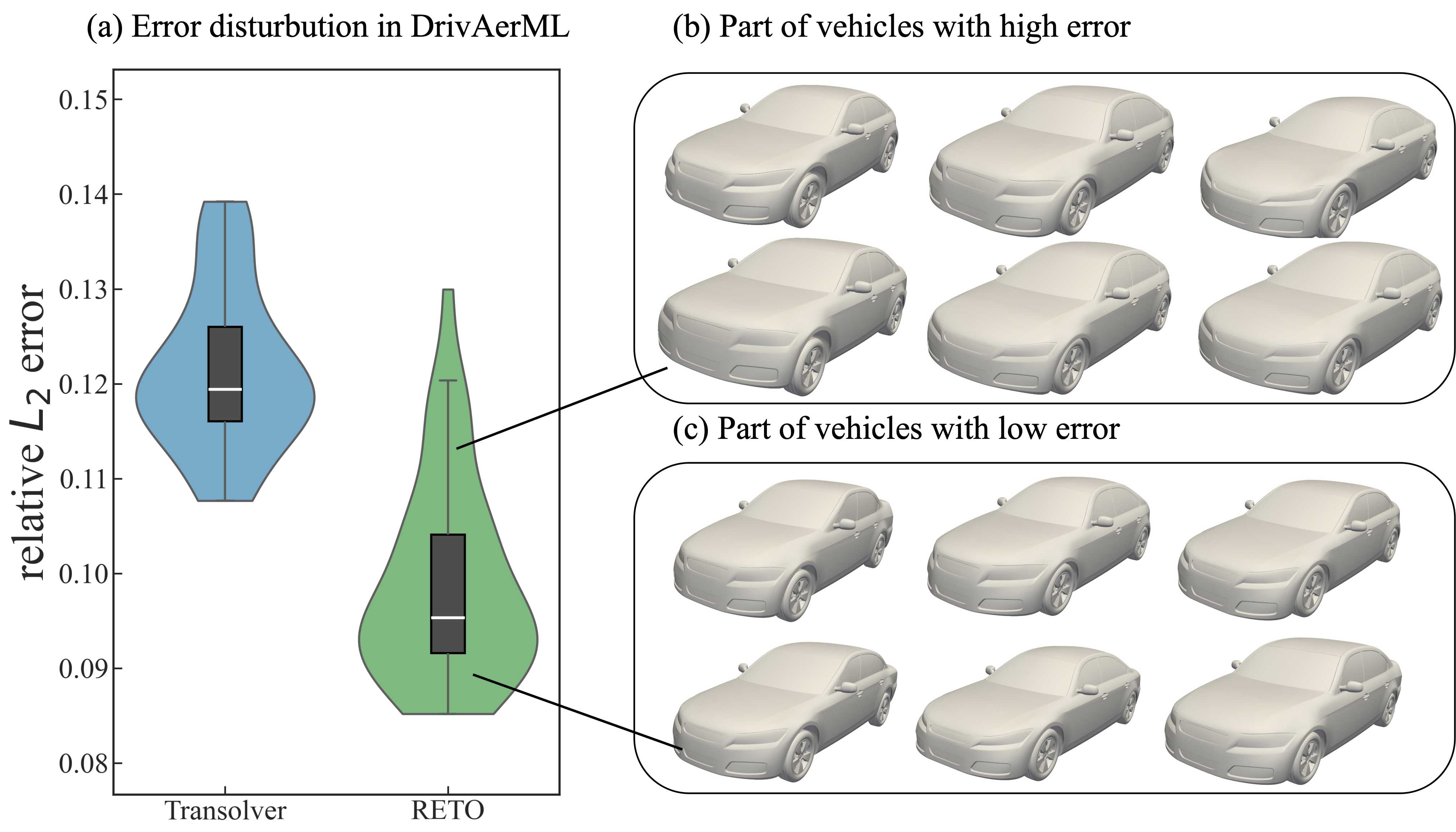}
    \caption{(a) The velocity error distribution of Transolver and RETO for $L_2$ error in DrivAerML Car. 
    The six vehicle types with the highest (b) and lowest (c) errors in the test set using the RETO model.}
    \label{fig:violin_DrivAerML}
\end{figure*}
In Fig.~\ref{fig:violin_DrivAerML} (a), we further assess the performance of the RETO model in comparison with Transolver on the DrivAerML dataset through the distribution of relative $L_2$ errors. It can be observed that RETO consistently achieves lower error values across the entire distribution. In particular, the interquartile range of RETO is noticeably more compact, indicating improved prediction stability. Additionally, the upper tail of the error distribution is significantly suppressed, suggesting that RETO is more effective in reducing large-error cases. These results highlight the superior generalization capability of RETO when applied to complex automotive geometries. Qualitatively, Fig.~\ref{fig:violin_DrivAerML} (b) and (c) illustrate representative vehicle geometries from the high-error and low-error regimes within the DrivAerML dataset \cite{ashton2024drivaerml}, respectively. All depicted samples belong to the Notchback family, highlighting that even within a consistent topological class, subtle variations in the 16 design parameters can significantly alter the complexity of the induced flow fields. The high-error instances in (b) correspond to specific parametric configurations where geometric modifications trigger more sensitive aerodynamic responses, such as intricate flow separation at the rear. Conversely, the low-error cases in (c) demonstrate the proficiency of RETO in capturing the fundamental physical patterns across diverse notchback profiles. The overall concentration of errors in the low-magnitude range confirms the model robustness in resolving the nonlinear mapping between high-dimensional geometric parameters and their corresponding aerodynamic fields.

\begin{figure*}
    \centering
    \includegraphics[width=0.7\textwidth]{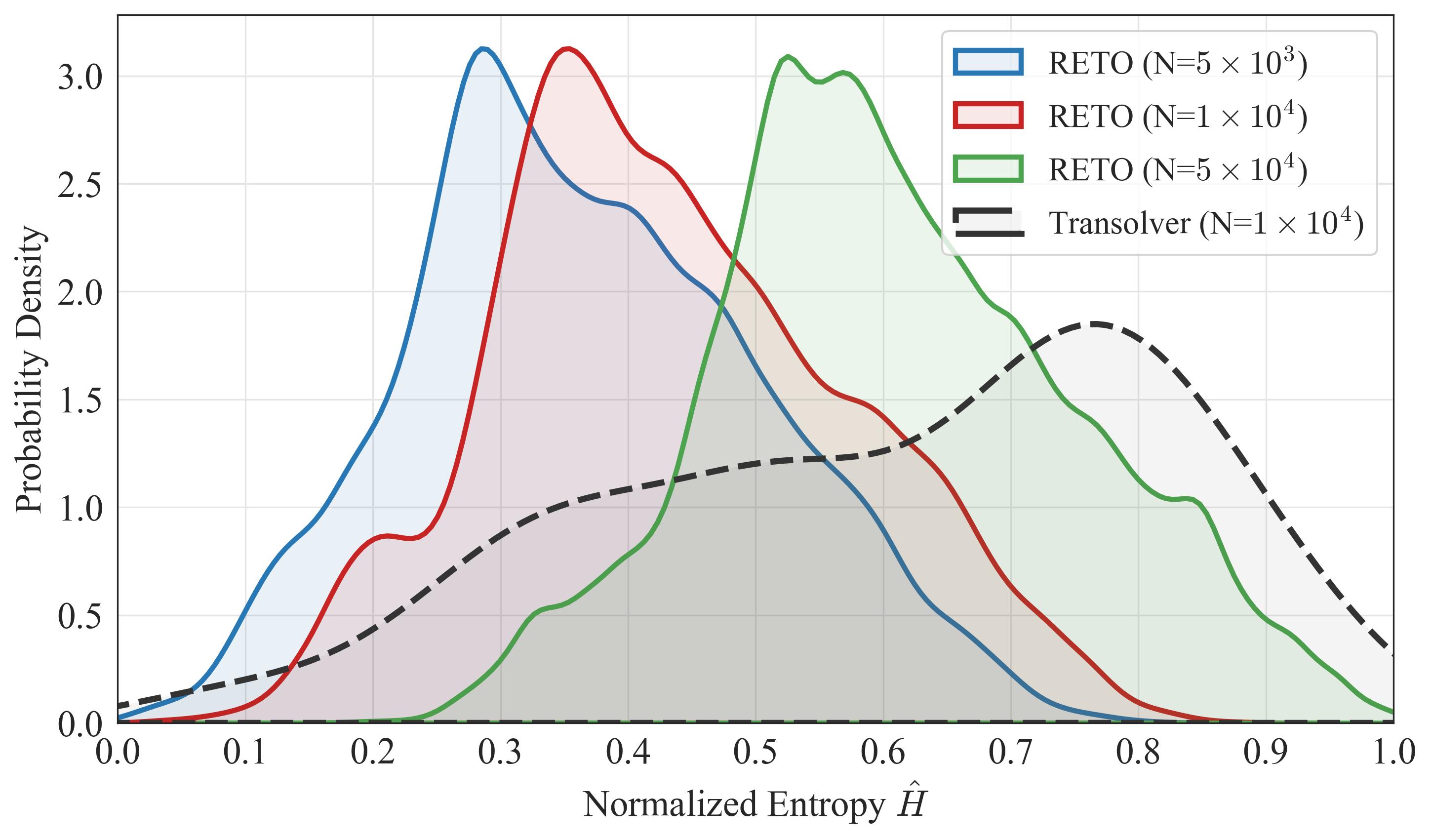}
    \caption{Comparative Analysis of Attention Entropy: RETO vs. Transolver}
    \label{fig:Entropy}
\end{figure*}

To quantify the focus and resolution-invariance of the attention mechanism, we define the normalized attention entropy \cite{2021Calibration}. For a given query point $x_i$, let $\mathcal{A}_i = \{A_{i,1}, A_{i,2}, \dots, A_{i,N}\}$ be the attention weights assigned to all $N$ key points. The Shannon entropy $H_i$ is calculated as:
\begin{equation}
H_i = -\sum_{j=1}^{N} A_{i,j} \ln (A_{i,j} + \epsilon)\,,
\end{equation}
and $\epsilon$ denotes a minimal regularization constant (typically set to $1 \times 10^{-12}$) to prevent numerical singularities in the presence of sparse attention weights.

To enable a consistent comparison across varying mesh resolutions, we normalize the entropy by the maximum possible value $\ln N$:
\begin{equation}
\hat{H}_i = \frac{H_i}{\ln N}\,,
\end{equation}
the normalized entropy $\hat{H} \in [0, 1]$ serves as a proxy for the sharpness of the attention kernel: $\hat{H} \to 0$ indicates a highly localized distribution where the model selectively attends to key geometric features, while $\hat{H} \to 1$ implies a diffuse, nearly uniform distribution. Based on this metric, the PDFs of $\hat{H}$ is evaluated to scrutinize the attention characteristics of the RETO model. As illustrated in Fig.\ref{fig:Entropy}, RETO exhibits a significantly sharper and more left-leaning distribution compared to the Transolver baseline across all tested resolutions. Specifically, at $N=10^4$, the peak of RETO entropy distribution is situated near $\hat{H} \approx 0.35$, whereas Transolver  distribution is considerably broader with a peak residing near $\hat{H} \approx 0.75$. This disparity indicates that RETO maintains a focused representation of spatial dependencies, whereas the baseline suffers from global attention diffusion, where the influence of local geometric gradients is diluted by excessive spatial averaging. This interpretation is consistent with prior studies showing that overly diffuse or over-globalized attention can dilute relevant local information and degrade prediction quality, whereas sharper attention distributions help preserve salient and task-relevant structures \cite{koh2025integrating}. Furthermore, our results demonstrate the remarkable stability of the entropy profile as the point cloud density increases from $5\times10^3$ to $5\times10^4$. Although a subtle rightward shift occurs—consistent with the increased information density and neighbor count in higher-resolution meshes—the PDF retains a consistent, highly structured unimodal shape. This robust behavior is primarily attributed to the integration of RoPE. By encoding relative spatial displacements into the rotary manifold, RoPE ensures that each query point maintains a consistent spatial attentional scope regardless of the total point count. Unlike absolute position encodings that often degrade when encountering unseen densities, RoPE provides the necessary inductive bias for the model to successfully extrapolate from sparse training data ($10^4$) to dense inference fields ($5\times10^4$) while preserving attention resolution and predictive integrity. 

\section{Conclusions}

In this study, we propose RETO, a novel neural solver designed to address the inherent limitations of standard transformer architectures in capturing the intricate spatial correlations required for automotive aerodynamics. The core of RETO predictive precision lies in its dual-stage spatial awareness mechanism, which synergistically combines global and relative positional information. By employing sinusoidal-cosine encodings, the model establishes a stable absolute coordinate frame that provides essential global structural referencing across the entire vehicle manifold. Building upon this, the integration of RoPE allows the self-attention mechanism to explicitly resolve relative spatial displacements through unitary rotations in the complex domain. This formulation is particularly critical for fluid dynamics: while absolute encodings often struggle with varying point densities, RoPE inherently preserves translation invariance and maintains a consistent receptive field. This enables RETO to resolve sharp localized physical gradients and complex flow separation zones—features that are typically attenuated by the excessive spatial smoothing inherent in the global averaging of baseline neural operators. 

The efficacy of RETO is rigorously validated across two distinct benchmarks, demonstrating consistent advancements over existing neural operators. On the ShapeNet dataset, RETO achieves a relative $L_2$ error of 0.063, whereas the error for the Transolver baseline is 0.075. This performance represents a 16\% improvement in predictive accuracy over Transolver, while the error for RegDGCNN remains significantly higher at 0.125. These results are further reinforced by evaluations on the industrial-grade DrivAerML dataset. In this high-fidelity context, RETO reduces the relative $L_2$ errors to 0.089 for surface pressure and 0.097 for velocity. In comparison, Transolver yields errors of 0.116 and 0.121 for the same metrics, indicating that RETO achieves substantial precision enhancements of 23\% and 19\% over Transolver in pressure and velocity fields, respectively. For comprehensive comparison, the errors for AB-UBT are recorded at 0.102 and 0.124, while RegDGCNN results in 0.235 and 0.312, respectively. Overall, these findings confirm that the proposed dual-stage spatial awareness mechanism enables RETO to maintain superior geometric generalization and localized precision across varying levels of geometric fidelity.

Future research will prioritize balancing this predictive fidelity with computational efficiency, specifically through the implementation of linear-complexity attention mechanisms and hierarchical operator structures to facilitate rapid, large-scale aerodynamic design iterations. In parallel, another important direction is the extension to unsteady flow modeling. We acknowledge that current mainstream open-source datasets for Automotive Aerodynamics only provide steady-state flow fields without transient information. It is crucial to systematically establish a dataset of time series of unsteady flow.

\section*{Acknowledgments}
The authors wish to extend their sincere gratitude to Keqi Ding and Hui Li from Tenfong Technology Co. Ltd., Zhou Jiang and Shi Yang from Chongqing University, and Ning Chang from Ningxia University for the insightful discussions and invaluable assistance throughout this study.

This work was supported by the National Natural Science Foundation of China (NSFC) (Grant Nos. 12588301 and 12302283); the National Key R\&D Program of China (Grant No. 2024YFF1500600); the NSFC Excellence Research Group Program for ‘Multiscale Problems in Nonlinear Mechanics’ (No. 12588201); the Shenzhen Science and Technology Program (Grant Nos.SYSPG20241211173725008,and KQTD20180411143441009); and the Department of Science and Technology of Guangdong Province (Grant Nos. 2019B21203001, 2020B1212030001 and 2023B1212060001). Additional support was provided by the Innovation Capability Support Program of Shaanxi (Program No. 2023-CX-TD-30) and the Center for Computational Science and Engineering of Southern University of Science and Technology.


\bibliographystyle{unsrt}
\bibliography{ref}   

@article{abinesh2014cfd,
  title={{CFD analysis of aerodynamic drag reduction and improve fuel economy}},
  author={Abinesh, J and Arunkumar, J},
  journal={International Journal of Mechanical Engineering and Robotics Research},
  volume={3},
  number={4},
  pages={430},
  year={2014},
  publisher={International Journal of Life Sciences Biotechnology and Pharma Research}
}

@techreport{bibra2022global,
  title = {{Global EV Outlook 2022: Securing Supplies for an Electric Future}},
  author = {Bibra, Ekta Meena and Connelly, Elizabeth and Dhir, Shobhan and Drtil, Michael and Henriot, Pauline and Hwang, Inchan and Le Marois, Jean‑Baptiste and McBain, Sarah and Paoli, Leonardo and Teter, Jacob},
  institution = {International Energy Agency},
  year = {2022},
}

@article{ang2023aerodynamic,
  title={{Aerodynamic optimisation of prototype FSAE vehicle through biomimetic approach}},
  author={Ang, Regina Jing Wen and Chong, Kok Hing and Sia, Charlie ChinVoon and Julaihi, Muhammad Rafiq Mirza},
  journal={Materials Today: Proceedings},
  volume={72},
  pages={2869--2874},
  year={2023},
  publisher={Elsevier}
}

@article{islam2017detailed,
  title={{A detailed statistical study of unsteady wake dynamics from automotive bluff bodies}},
  author={Islam, Asiful and Gaylard, Adrian and Thornber, Ben},
  journal={Journal of Wind Engineering and Industrial Aerodynamics},
  volume={171},
  pages={161--177},
  year={2017},
  publisher={Elsevier}
}

@article{zhang2020experimental,
  title={{Experimental investigation on wake flow structures of Motor Industry Research Association square-back model}},
  author={Zhang, Yingchao and Wang, Ruidong and Yang, Chao and Wang, Zijie and Zhang, Zhe},
  journal={Advances in Mechanical Engineering},
  volume={12},
  number={6},
  pages={1687814020932313},
  year={2020},
  publisher={SAGE Publications Sage UK: London, England}
}

@article{2013Interference,
  title={{Interference effects of cooling airflows on a generic car body}},
  author={ Bder, Dirk  and  Indinger, Thomas  and  Adams, Nikolaus A.  and  Unterlechner, Peter  and  Wickern, Gerhard },
  journal={Journal of Wind Engineering and Industrial Aerodynamics},
  volume={119},
  number={8},
  pages={146-157},
  year={2013},
}

@article{2023Automotive,
  title={{Automotive Aerodynamic Drag and Lift Analysis using Computational Fluid Dynamics Software: A Review}},
  author={Madha, Jowad Md  and  Nawar, Anika  and  Rahman, Md Mizanur },
  journal={Proceedings of the International Conference on Industrial Engineering and Operations Management},
  year={2023},
}

@book{anderson1995computational,
  title={{Computational fluid dynamics}},
  author={Anderson, John David and Wendt, John and others},
  volume={206},
  year={1995},
  publisher={Springer}
}

@article{pope2001turbulent,
  title={{Turbulent flows}},
  author={Pope, S.B.},
  journal={Cambridge University Press, Cambridge},
  year={2000}
}

@book{ferziger2019computational,
  title={{Computational methods for fluid dynamics}},
  author={Ferziger, Joel H and Peri{\'c}, Milovan and Street, Robert L},
  year={2019},
  publisher={springer}
}

@article{aultman2022evaluation,
  title={{Evaluation of CFD methodologies for prediction of flows around simplified and complex automotive models}},
  author={Aultman, Matthew and Wang, Zhenyu and Auza-Gutierrez, Rodrigo and Duan, Lian},
  journal={Computers \& Fluids},
  volume={236},
  pages={105297},
  year={2022},
  publisher={Elsevier}
}

@book{smith1985numerical,
  title={{Numerical solution of partial differential equations: finite difference methods}},
  author={Smith, Gordon D},
  year={1985},
  publisher={Oxford university press}
}

@book{hughes2003finite,
  title={{The finite element method: linear static and dynamic finite element analysis}},
  author={Hughes, Thomas JR},
  year={2003},
  publisher={Courier Corporation}
}

@article{eymard2000finite,
  title={{Finite volume methods}},
  author={Eymard, Robert and Gallou{\"e}t, Thierry and Herbin, Rapha{\`e}le},
  journal={Handbook of numerical analysis},
  volume={7},
  pages={713--1018},
  year={2000},
  publisher={Elsevier}
}

@article{chen1998lattice,
  title={{Lattice Boltzmann method for fluid flows}},
  author={Chen, Shiyi and Doolen, Gary D},
  journal={Annual review of fluid mechanics},
  volume={30},
  number={1},
  pages={329--364},
  year={1998},
  publisher={Annual Reviews 4139 El Camino Way, PO Box 10139, Palo Alto, CA 94303-0139, USA}
}

@article{lee2022development,
  title={{Development of high-fidelity numerical methodology for prediction of vehicle interior noise due to external flow disturbances using LES and vibroacoustic techniques}},
  author={Lee, Songjune and Lee, Sang-heon and Cheong, Cheolung},
  journal={Applied Sciences},
  volume={12},
  number={13},
  pages={6345},
  year={2022},
  publisher={MDPI}
}

@article{lu2021learning,
  title={{Learning nonlinear operators via {DeepONet} based on the universal approximation theorem of operators}},
  author={Lu, Lu and Jin, Pengzhan and Pang, Guofei and Zhang, Zhongqiang and Karniadakis, George Em},
  journal={Nature machine intelligence},
  volume={3},
  number={3},
  pages={218--229},
  year={2021},
  publisher={Nature Publishing Group UK London}
}

@article{garcia2023cnn,
  title={{CNN-based flow field prediction for bus aerodynamics analysis}},
  author={Garcia-Fernandez, Roberto and Portal-Porras, Koldo and Irigaray, Oscar and Ansa, Zugatz and Fernandez-Gamiz, Unai},
  journal={Scientific Reports},
  volume={13},
  number={1},
  pages={21213},
  year={2023},
  publisher={Nature Publishing Group UK London}
}

@article{elrefaie2024drivaernet++,
  title={{Drivaernet++: A large-scale multimodal car dataset with computational fluid dynamics simulations and deep learning benchmarks}},
  author={Elrefaie, Mohamed and Morar, Florin and Dai, Angela and Ahmed, Faez},
  journal={Advances in Neural Information Processing Systems},
  volume={37},
  pages={499--536},
  year={2024}
}

@article{vaswani2017attention,
  title={{Attention is all you need}},
  author={Vaswani, Ashish and Shazeer, Noam and Parmar, Niki and Uszkoreit, Jakob and Jones, Llion and Gomez, Aidan N and Kaiser, {\L}ukasz and Polosukhin, Illia},
  journal={Advances in neural information processing systems},
  volume={30},
  year={2017}
}

@article{qi2017pointnet++,
  title={{Pointnet++: Deep hierarchical feature learning on point sets in a metric space}},
  author={Qi, Charles Ruizhongtai and Yi, Li and Su, Hao and Guibas, Leonidas J},
  journal={Advances in neural information processing systems},
  volume={30},
  year={2017}
}

@article{jiang2023transcfd,
  title={{TransCFD: A transformer-based decoder for flow field prediction}},
  author={Jiang, Jundou and Li, Guanxiong and Jiang, Yi and Zhang, Laiping and Deng, Xiaogang},
  journal={Engineering Applications of Artificial Intelligence},
  volume={123},
  pages={106340},
  year={2023},
  publisher={Elsevier}
}

@article{yang2025spatially,
  title={{Spatially-aware transformer operator for real-time aerodynamic evaluations of arbitrary three-dimensional vehicles}},
  author={Yang, Huiyu and Gu, Jianghang and Chen, Yuntian and Bin, Yuanwei and Wang, Jianchun and Chen, Shiyi},
  journal={Journal of Computational Physics},
  pages={114456},
  year={2025},
  publisher={Elsevier}
}

@article{li2022transformer,
  title={{Transformer for partial differential equations' operator learning}},
  author={Li, Zijie and Meidani, Kazem and Farimani, Amir Barati},
  journal={arXiv preprint arXiv:2205.13671},
  year={2022}
}

@article{alkin2025ab,
  title={{AB-UPT: Scaling Neural CFD Surrogates for High-Fidelity Automotive Aerodynamics Simulations via Anchored-Branched Universal Physics Transformers}},
  author={Alkin, Benedikt and Bleeker, Maurits and Kurle, Richard and Kronlachner, Tobias and Sonnleitner, Reinhard and Dorfer, Matthias and Brandstetter, Johannes},
  journal={arXiv preprint arXiv:2502.09692},
  year={2025}
}

@article{ashton2024drivaerml,
  title={{DrivAerML: High-fidelity computational fluid dynamics dataset for road-car external aerodynamics}},
  author={Ashton, Neil and Mockett, Charles and Fuchs, Marian and Fliessbach, Louis and Hetmann, Hendrik and Knacke, Thilo and Schonwald, Norbert and Skaperdas, Vangelis and Fotiadis, Grigoris and Walle, Astrid and others},
  journal={arXiv preprint arXiv:2408.11969},
  year={2024}
}

@article{wu2024transolver,
  title={{Transolver: A fast transformer solver for PDEs on general geometries}},
  author={Wu, Haixu and Luo, Huakun and Wang, Haowen and Wang, Jianmin and Long, Mingsheng},
  journal={arXiv preprint arXiv:2402.02366},
  year={2024}
}

@article{bhatnagar2019prediction,
  title={{Prediction of aerodynamic flow fields using convolutional neural networks}},
  author={Bhatnagar, Saakaar and Afshar, Yaser and Pan, Shaowu and Duraisamy, Karthik and Kaushik, Shailendra},
  journal={Computational Mechanics},
  volume={64},
  number={2},
  pages={525--545},
  year={2019},
  publisher={Springer}
}

@article{wang2019dynamic,
  title={{Dynamic graph CNN for learning on point clouds}},
  author={Wang, Yue and Sun, Yongbin and Liu, Ziwei and Sarma, Sanjay E and Bronstein, Michael M and Solomon, Justin M},
  journal={ACM Transactions on Graphics (tog)},
  volume={38},
  number={5},
  pages={1--12},
  year={2019},
  publisher={Acm New York, NY, USA}
}

@article{elrefaie2025drivaernet,
  title={{Drivaernet: A parametric car dataset for data-driven aerodynamic design and prediction}},
  author={Elrefaie, Mohamed and Dai, Angela and Ahmed, Faez},
  journal={Journal of Mechanical Design},
  volume={147},
  number={4},
  pages={041712},
  year={2025},
  publisher={American Society of Mechanical Engineers}
}

@inproceedings{chen20213d,
  title={{3D Flow Field Estimation around a Vehicle Using Convolutional Neural Networks}},
  author={Chen, Fangge and Akasaka, Kei},
  booktitle={BMVC},
  pages={396},
  year={2021}
}

@article{lu2019deeponet,
  title={{Deeponet: Learning nonlinear operators for identifying differential equations based on the universal approximation theorem of operators}},
  author={Lu, Lu and Jin, Pengzhan and Karniadakis, George Em},
  journal={arXiv preprint arXiv:1910.03193},
  year={2019}
}

@article{kovachki2023neural,
  title={{Neural operator: Learning maps between function spaces with applications to PDEs}},
  author={Kovachki, Nikola and Li, Zongyi and Liu, Burigede and Azizzadenesheli, Kamyar and Bhattacharya, Kaushik and Stuart, Andrew and Anandkumar, Anima},
  journal={Journal of Machine Learning Research},
  volume={24},
  number={89},
  pages={1--97},
  year={2023}
}

@article{li2020fourier,
  title={{Fourier neural operator for parametric partial differential equations}},
  author={Li, Zongyi and Kovachki, Nikola and Azizzadenesheli, Kamyar and Liu, Burigede and Bhattacharya, Kaushik and Stuart, Andrew and Anandkumar, Anima},
  journal={arXiv preprint arXiv:2010.08895},
  year={2020}
}

@article{tran2021factorized,
  title={{Factorized Fourier neural operators}},
  author={Tran, Alasdair and Mathews, Alexander and Xie, Lexing and Ong, Cheng Soon},
  journal={arXiv preprint arXiv:2111.13802},
  year={2021}
}

@article{touvron2023llama,
  title={{Llama: Open and efficient foundation language models}},
  author={Touvron, Hugo and Lavril, Thibaut and Izacard, Gautier and Martinet, Xavier and Lachaux, Marie-Anne and Lacroix, Timoth{\'e}e and Rozi{\`e}re, Baptiste and Goyal, Naman and Hambro, Eric and Azhar, Faisal and others},
  journal={arXiv preprint arXiv:2302.13971},
  year={2023}
}

@article{wu2022flowformer,
  title={{Flowformer: Linearizing transformers with conservation flows}},
  author={Wu, Haixu and Wu, Jialong and Xu, Jiehui and Wang, Jianmin and Long, Mingsheng},
  journal={arXiv preprint arXiv:2202.06258},
  year={2022}
}

@article{ahmed1984some,
  title={{Some salient features of the time-averaged ground vehicle wake}},
  author={Ahmed, Syed R and Ramm, G and Faltin, Gunter},
  journal={SAE transactions},
  pages={473--503},
  year={1984},
  publisher={JSTOR}
}

@article{schmitt2007boussinesq,
  title={{About Boussinesq's turbulent viscosity hypothesis: historical remarks and a direct evaluation of its validity}},
  author={Schmitt, Fran{\c{c}}ois G},
  journal={Comptes Rendus M{\'e}canique},
  volume={335},
  number={9-10},
  pages={617--627},
  year={2007},
  publisher={Elsevier}
}

@article{swanson2021solving,
  title={{Solving Two-Equation Turbulence Models With a Perspective on Solving Transport Equations}},
  author={Swanson, R Charles},
  year={2021}
}

@book{wilcox1998turbulence,
  title={{Turbulence modeling for CFD}},
  author={Wilcox, David C and others},
  volume={2},
  year={1998},
  publisher={DCW industries La Canada, CA}
}

@article{menter2003ten,
  title={{Ten years of industrial experience with the SST turbulence model}},
  author={Menter, Florian R and Kuntz, Martin and Langtry, Robin and others},
  journal={Turbulence, heat and mass transfer},
  volume={4},
  number={1},
  pages={625--632},
  year={2003}
}

@article{2024DrivAerNet,
  title={{DrivAerNet: A Parametric Car Dataset for Data-Driven Aerodynamic Design and Prediction}},
  author={Elrefaie, Mohamed  and  Dai, Angela  and  Ahmed, Faez},
  year={2024},
}

@article{chang2015shapenet,
  title={{Shapenet: An information-rich 3D model repository}},
  author={Chang, Angel X and Funkhouser, Thomas and Guibas, Leonidas and Hanrahan, Pat and Huang, Qixing and Li, Zimo and Savarese, Silvio and Savva, Manolis and Song, Shuran and Su, Hao and others},
  journal={arXiv preprint arXiv:1512.03012},
  year={2015}
}

@article{spalart2009detached,
  title={{Detached-eddy simulation}},
  author={Spalart, Philippe R},
  journal={Annual review of fluid mechanics},
  volume={41},
  number={1},
  pages={181--202},
  year={2009},
  publisher={Annual Reviews}
}

@article{hupertz2021aerodynamics,
  title={{On the aerodynamics of the notchback open cooling DrivAer: A detailed investigation of wind tunnel data for improved correlation and reference}},
  author={Hupertz, Burkhard and Chalupa, Karel and Krueger, Lothar and Howard, Kevin and Glueck, Hans-Dieter and Lewington, Neil and Chang, Jin-Hyuck and Shin, Yong-su},
  journal={SAE International Journal of Advances and Current Practices in Mobility},
  volume={3},
  number={2021-01-0958},
  pages={1726--1747},
  year={2021}
}

@article{su2024roformer,
  title={{Roformer: Enhanced transformer with rotary position embedding}},
  author={Su, Jianlin and Ahmed, Murtadha and Lu, Yu and Pan, Shengfeng and Bo, Wen and Liu, Yunfeng},
  journal={Neurocomputing},
  volume={568},
  pages={127063},
  year={2024},
  publisher={Elsevier}
}

@article{wilcox2008formulation,
  title={{Formulation of the KW turbulence model revisited}},
  author={Wilcox, David C},
  journal={AIAA journal},
  volume={46},
  number={11},
  pages={2823--2838},
  year={2008}
}

@techreport{menter1992improved,
  title={{Improved two-equation k-omega turbulence models for aerodynamic flows}},
  author={Menter, Florian R},
  year={1992}
}

@incollection{piomelli1996large,
  title={{Large-eddy simulations: theory and applications}},
  author={Piomelli, Ugo and Chasnov, Jeffrey Robert},
  booktitle={Turbulence and Transition Modelling: Lecture Notes from the ERCOFTAC/IUTAM Summerschool held in Stockholm, 12--20 June, 1995},
  pages={269--336},
  year={1996},
  publisher={Springer}
}

@article{krizhevsky2012imagenet,
  title={{Imagenet classification with deep convolutional neural networks}},
  author={Krizhevsky, Alex and Sutskever, Ilya and Hinton, Geoffrey E},
  journal={Advances in neural information processing systems},
  volume={25},
  year={2012}
}

@article{lecun2002gradient,
  title={{Gradient-based learning applied to document recognition}},
  author={LeCun, Yann and Bottou, L{\'e}on and Bengio, Yoshua and Haffner, Patrick},
  journal={Proceedings of the IEEE},
  volume={86},
  number={11},
  pages={2278--2324},
  year={2002},
  publisher={Ieee}
}

@article{luo2025transolver++,
  title={{Transolver++: An accurate neural solver for PDEs on million-scale geometries}},
  author={Luo, Huakun and Wu, Haixu and Zhou, Hang and Xing, Lanxiang and Di, Yichen and Wang, Jianmin and Long, Mingsheng},
  journal={arXiv preprint arXiv:2502.02414},
  year={2025}
}

@article{du2025spatiotemporal,
  title={{Spatiotemporal Field Generation Based on Hybrid Mamba-Transformer with Physics-informed Fine-tuning}},
  author={Du, Peimian and Liu, Jiabin and Jin, Xiaowei and Zuo, Wangmeng and Li, Hui},
  journal={arXiv preprint arXiv:2505.11578},
  year={2025}
}

@article{curvo2025mspt,
  title={{MSPT: Efficient Large-Scale Physical Modeling via Parallelized Multi-Scale Attention}},
  author={Curvo, Pedro MP and van de Meent, Jan-Willem and Zhdanov, Maksim},
  journal={arXiv preprint arXiv:2512.01738},
  year={2025}
}

@article{adams2025geotransolver,
  title={{GeoTransolver: Learning Physics on Irregular Domains Using Multi-scale Geometry Aware Physics Attention Transformer}},
  author={Adams, Corey and Ranade, Rishikesh and Cherukuri, Ram and Choudhry, Sanjay},
  journal={arXiv preprint arXiv:2512.20399},
  year={2025}
}

@book{wasserman2004all,
  title={{All of Statistics: A Concise Course in Statistical Inference}},
  author={Wasserman, Larry},
  year={2004},
  publisher={Springer}
}

@book{2006Numerical,
  title={{Numerical Mathematics}},
  author={Quarteroni, Alfio  and  Sacco, Riccardo  and  Saleri, Fausto},
  publisher={Numerical Mathematics},
  year={2006},
}

@article{1990The,
  title={{The numerical computation of turbulent flows}},
  author={ Launder, Brian E  and  Spalding, D B },
  journal={Computer Methods in Applied Mechanics and Engineering},
  year={1990},
}

@article{2014Adam,
  title={{Adam: A Method for Stochastic Optimization}},
  author={Kingma, Diederik  and  Ba, Jimmy},
  journal={Computer Science},
  year={2014},
}

@article{zhou2026transolver,
  title={{Transolver-3: Scaling Up Transformer Solvers to Industrial-Scale Geometries}},
  author={Zhou, Hang and Wu, Haixu and Shangguan, Haonan and Ma, Yuezhou and Weng, Huikun and Wang, Jianmin and Long, Mingsheng},
  journal={arXiv preprint arXiv:2602.04940},
  year={2026}
}

@inproceedings{2021Calibration,
  title={{Calibration, Entropy Rates, and Memory in Language Models}},
  author={Braverman, Mark  and  Chen, Xinyi  and  Kakade, Sham  and  Narasimhan, Karthik  and  Zhang, Cyril  and  Zhang, Yi},
  booktitle={37th International Conference on Machine Learning: ICML 2020, Online, 13-18 July 2020, Part 2 of 15},
  year={2021},
}

@article{2014Neural,
  title={{Neural Machine Translation by Jointly Learning to Align and Translate}},
  author={Bahdanau, Dzmitry  and  Cho, Kyunghyun  and  Bengio, Yoshua },
  journal={Computer Science},
  year={2014},
}

@article{2015Effective,
  title={{Effective Approaches to Attention-based Neural Machine Translation}},
  author={Luong, Minh Thang  and  Pham, Hieu  and  Manning, Christopher D},
  journal={Computer ence},
  year={2015},
}

@inproceedings{liu2025dragsolver,
  title={{DragSolver: A multi-scale transformer for real-world automotive drag coefficient estimation}},
  author={Liu, Ye and Chen, Yuntian},
  booktitle={Forty-second International Conference on Machine Learning},
  year={2025}
}

@article{gu5601950geoformer,
  title={{GeoFormer: Mesh-Free Geometry-to-Flow Alignment Framework for Real-Time Aerodynamics on Non-Watertight CAD}},
  author={Gu, Jianghang and Chen, Yuntian and Bin, Yuanwei and Chen, Shiyi},
  journal={Available at SSRN 5601950}
}

@inproceedings{tsai2019transformer,
  title={{Transformer dissection: An unified understanding for transformer’s attention via the lens of kernel}},
  author={Tsai, Yao-Hung Hubert and Bai, Shaojie and Yamada, Makoto and Morency, Louis-Philippe and Salakhutdinov, Ruslan},
  booktitle={Proceedings of the 2019 conference on empirical methods in natural language processing and the 9th international joint conference on natural language processing (EMNLP-IJCNLP)},
  pages={4344--4353},
  year={2019}
}

@inproceedings{koh2025integrating,
  title={Integrating Locality-Aware Attention with Transformers for General Geometry PDEs},
  author={Koh, Minsu and Park, Beom-Chul and Kong, Heejo and Lee, Seong-Whan},
  booktitle={2025 International Joint Conference on Neural Networks (IJCNN)},
  pages={1--8},
  year={2025},
  organization={IEEE}
}

@article{elrefaie2025carbench,
  title={CarBench: A Comprehensive Benchmark for Neural Surrogates on High-Fidelity 3D Car Aerodynamics},
  author={Elrefaie, Mohamed and Shu, Dule and Klenk, Matt and Ahmed, Faez},
  journal={arXiv preprint arXiv:2512.07847},
  year={2025}
}

@inproceedings{Liu2026EMOS,
  title     = {EMOS: Efficient Multi-Output Aerodynamic Surrogates for Rapid Vehicle Design Iteration},
  author    = {Liu, Y. and Chen, Y. and Bin, Y. and Chen, S.},
  booktitle = {Proceedings of the IEEE International Conference on Multimedia and Expo (ICME)},
  year      = {2026}
}

@inproceedings{Liu2026AeroAgent,
  title     = {AeroAgent: A Vision–Physics–Decision Framework for Aerodynamic Vehicle Design},
  author    = {Liu, Y. and Liu, S. and Yang, H. and Gu, J. and Fan, W. and Yang, Z. and Wang, D. and Chen, S. and Jiang, Z. and Bin, Y. and Chen, S. and Chen, Y.},
  booktitle = {Proceedings of the IEEE/CVF Conference on Computer Vision and Pattern Recognition (CVPR)},
  year      = {2026}
}

@article{huang2024neural,
  title={Neural network-augmented SED-SL modeling of turbulent flows over airfoils},
  author={Huang, Wenxiao and Liu, Yilang and Bi, Weitao and Gao, Yizhuo and Chen, Jun},
  journal={Acta Mechanica Sinica},
  volume={40},
  number={3},
  pages={323517},
  year={2024},
  publisher={Springer}
}

@article{shan2024modeling,
  title={Modeling Reynolds stress anisotropy invariants via machine learning},
  author={Shan, Xianglin and Sun, Xuxiang and Cao, Wenbo and Zhang, Weiwei and Xia, Zhenhua},
  journal={Acta Mechanica Sinica},
  volume={40},
  number={6},
  pages={323629},
  year={2024},
  publisher={Springer}
}

@article{hao2025large,
  title={Large eddy simulation of low-Reynolds-number flow past the SD7003 airfoil with an improved high-precision IPDG method},
  author={Hao, Shixi and Zhao, Ming and Ding, Qiushi and Xiao, Jiabing and Chen, Yanan and Liu, Wei and Li, Xiaojian and Liu, Zhengxian},
  journal={Acta Mechanica Sinica},
  volume={41},
  number={2},
  pages={323637},
  year={2025},
  publisher={Springer}
}

@article{liu2026reconstructing,
  title={Reconstructing flow fields from sparse measurements using a convolutional autoencoder integrated with an Informer model},
  author={Liu, Yulu and Long, Jiakun and Wang, Bofu and Chang, Tienchong and Qiu, Xiang},
  journal={Acta Mechanica Sinica},
  volume={42},
  number={7},
  pages={725013},
  year={2026},
  publisher={Springer}
}

@article{bai2025towards,
  title={Towards the future of physics-and data-guided AI frameworks in computational mechanics},
  author={Bai, Jinshuai and Wang, Yizheng and Jeong, Hyogu and Chu, Shiyuan and Wang, Qingxia and Alzubaidi, Laith and Zhuang, Xiaoying and Rabczuk, Timon and Xie, Yi Min and Feng, Xi-Qiao and others},
  journal={Acta Mechanica Sinica},
  volume={41},
  number={7},
  pages={225340},
  year={2025},
  publisher={Springer}
}
\end{multicols}
\end{document}